\documentclass{mn2e}
\usepackage{epsfig,natbib2,natbibmnfix}
\usepackage{amsmath}
\usepackage{mathrsfs}
\usepackage{subfigure}
\usepackage{url}
\usepackage[dvips]{color}
\usepackage{multicol}

\newcommand{\be}{\begin{equation}}
\newcommand{\ee}{\end{equation}}

\def\ltsima{$\; \buildrel < \over \sim \;$}
\def\simlt{\lower.5ex\hbox{\ltsima}}
\def\gtsima{$\; \buildrel > \over \sim \;$}
\def\simgt{\lower.5ex\hbox{\gtsima}}

\def\del#1{{}}

\def\msun{{\,{\rm M}_\odot}}
\def\tsun{{\,{\rm t}_\odot}}

\def\unabla{{\bf \nabla}}
\def\uv{{\bf v}}

\def\ur{{\bf r}}

\def\wij{W_{ij}}

\def\dthr{\mathrm{d}^3{ r}}

\title[Thermal instabilities in galactic coronae]
{Thermal instabilities in cooling galactic coronae: fuelling star formation in galactic discs}

\author[Alexander Hobbs, Justin Read, Chris Power, David Cole]
       {\parbox{18cm}{Alexander Hobbs$^{1}$, Justin Read$^{1,3}$, Chris Power$^{2}$, David Cole$^{3}$}\vspace{0.3cm}\\
\noindent $^{1}$Institute for Astronomy, ETH Zurich $^{2}$University of Western Australia $^{3}$Dept. of Physics \& Astronomy, University of Leicester}

\begin{document}

\maketitle

\begin{abstract}
We investigate the means by which cold gas can accrete onto Milky Way mass galaxies from a hot corona of gas, using a new smoothed particle hydrodynamics code, `SPHS'. We find that the `cold clumps' seen in many classic SPH simulations in the literature are not present in our SPHS simulations. Instead, cold gas condenses from the halo along filaments that form at the intersection of supernovae-driven bubbles from previous phases of star formation. This {\it positive feedback} feeds cold gas to the galactic disc directly, fuelling further star formation. The resulting galaxies in the SPH and SPHS simulations differ greatly in their morphology, gas phase diagrams, and stellar content. We show that the classic SPH cold clumps owe to a numerical thermal instability caused by an inability for cold gas to mix in the hot halo. The improved treatment of mixing in SPHS suppresses this instability leading to a dramatically different physical outcome. In our highest resolution SPHS simulation, we find that the cold filaments break up into bound clumps that form stars. The filaments are overdense by a factor of 10-100 compared to the surrounding gas, suggesting that the fragmentation results from a physical non-linear instability driven by the overdensity. This `fragmenting filament' mode of disc growth has important implications for galaxy formation, in particular the role of star formation in bringing cold gas into disc galaxies. 
\end{abstract}

\renewcommand{\thefootnote}{\fnsymbol{footnote}}
\footnotetext[1]{E-mail: {\tt ahobbs@phys.ethz.ch}}

\section{Introduction}

The star formation rate (SFR) of the Universe has been rapidly falling since a redshift of $\sim 2-3$ \citep[e.g.,][]{LillyEtal1996, MadauEtal1998, HippeleinEtal2003}, dominated by the most massive galaxies and galaxy groups. By contrast, the Milky Way -- similarly to other disc galaxies -- has been forming stars at a near-continuous rate for the past $\sim 8$\,Gyr since $z \sim 1$ \citep[e.g.,][]{NohScalo1990, Rocha-PintoEtal2000}. The observed SFR of $\sim 1-3 \msun$ yr$^{-1}$ is hard to explain given the amount of cold gas present in the disc today; the Milky Way must have continuously accreted cold gas at a rate of $\simgt 1\,\msun$ yr$^{-1}$ over this time \citep{FraternaliTomassetti2012}. Merger-driven accretion appears to account for just $\sim 0.1 \msun $ yr$^{-1}$ \citep{SancisiEtal2008}, and there is to date no evidence of a low redshift `cold flow' accretion mode \citep[e.g.][]{StewartEtal2011}. This has led to two main solutions in the literature. The first is recycling of gas from existing stars in the disc through stellar winds \citep[e.g.,][]{Roberts1963, Sandage1986, KennicuttEtal1994}. \cite{LeitnerKravtsov2011} estimate that this recycled gas could contribute at least half of the global SFR for a galaxy of Milky Way mass at low redshift ($z \simlt 0.5$). The second is accretion from a massive hot halo, or corona, of gas surrounding the Galaxy. Such a hot halo has not yet been directly observed, but several indirect lines of evidence exist: observations of X-ray emitting gas \citep{GuptaEtal2012}; absorption along sight lines to quasars \citep[e.g.,][]{WilliamsEtal2005, FangEtal2006, KacprzakEtal2008}; pulsar dispersion measures \citep[e.g.,][]{2010ApJ...714..320A, GaenslerEtal2008}; a significant Galactic baryon deficiency when compared to the universal baryon fraction \citep[e.g.,][]{FukugitaPeebles2006, NicastroEtal2008}; and evidence of ram pressure stripping of the Magellanic stream \citep{MastropietroEtal2005} and other nearby dwarf galaxies \citep{GrcevichPutman2009}. These studies give a hot halo mass of $> 5 \times 10^9 - 1.5 \times 10^{10} \msun$ assuming a \cite{NavarroFrenkWhite1996} (NFW) profile or $\simgt 4 \times 10^{10}$ assuming a flattened power-law profile \citep{AndersonBregman2011}. Similar results are seen in other disc galaxies \citep[e.g.,][]{2005RSPTA.363.2693R, MoEtal2005, SancisiEtal2008, AndersonBregman2011}.


While it is likely that hot gaseous coronae surround disc galaxies, it is not clear \emph{how} the gas cools and condenses out of these coronae and onto the disc to form stars. One particular mode of cold gas supply that has been seen in a number of numerical simulations \citep{Sommer-Larsen2006, KaufmannEtal2006, KaufmannEtal2007, KaufmannEtal2009, PutmanEtal2009} but was initially proposed by \cite{Nulsen1986}, is that of direct cooling from the halo via thermal instability. However, doubt has been cast on this picture by \cite{MalagoliEtal1987} \& \cite{BinneyEtal2009} who show that hot haloes are linearly stable to density perturbations, making direct cooling unlikely. \cite{JoungEtal2012} extend this treatment to the non-linear regime with dedicated numerical simulations, finding that while a runaway process of cooling and collapse is possible, it can only occur for overdensities of $\simgt 10-20$ with respect to the local background density. These results suggest that if gas is to cool from the hot coronae then some mechanism is required to \emph{excite} a thermal instability. \cite{MarinacciEtal2011} suggest a mechanism whereby a galactic fountain seeds metal-rich gas into the metal-poor hot haloes, giving rise to a thermal instability that causes cold gas to rain down onto the disc in the form of $\sim 10^5 \msun$ clouds \citep[see also][]{FraternaliBinney2008}. This model provides an excellent fit to both the kinematics and spatial distribution of warm HI gas in the Milky Way \citep{MarascoEtal2012}, although it cannot account for the high-velocity clouds (HVCs) \citep[e.g.,][]{SembachEtal2003, TrippEtal2003, CollinsEtal2005}. Alternative models include cooling stripped gas from dwarf galaxies or warm clouds at the disc-corona interface \citep[e.g.,][]{PutmanEtal2009, HeitschPutman2009, Peek2009}.

The `direct cooling' mode mentioned above is seen regularly in `classic'\footnote{for a careful definition of `classic' SPH see Section \ref{sec:classic}.} smoothed particle hydrodynamics (SPH) simulations \citep[e.g.,][]{Sommer-Larsen2006, KaufmannEtal2006, KaufmannEtal2007, KaufmannEtal2009, PutmanEtal2009}, where a thermal instability leads to a sudden and widespread condensation of gas from the halo in the form of cold, dense clumps. If correct, no special mechanism would be required to explain the continued star formation of disc galaxies over the past $\sim 8$\,Gyrs. However, `classic' SPH is known to exhibit an artificial surface tension that inhibits mixing of different gaseous phases, leading to poor performance on a variety of hydrodynamic test problems \citep{AgertzEtal2007, 2008JCoPh.22710040P, 2008MNRAS.387..427W, ReadEtal2010}. In recent work, some of the present authors have developed a new `flavour' of SPH -- SPHS -- that solves these problems, giving excellent performance and numerical convergence on a wide range of test problems \citep{2012MNRAS.tmp.2941R}. In this paper, we use SPHS to revisit the problem of thermal instabilities in hot gaseous coronae. Our primary goals are to determine whether the instabilities seen in the classic SPH simulations are physical or numerical; and under what circumstances cold gas can condense out of a hot halo to fuel star formation in disc galaxies. The former goal is further motivated by the absence of such cold clumps in the hot haloes of galaxy formation simulations using adaptive mesh refinement (AMR) codes \citep[e.g.,][]{AgertzEtal2009} or the recent moving Voronoi mesh code AREPO \citep{VogelsbergerEtal2011}.

This paper is organised as follows. In Section \ref{sec:method}, we briefly review the SPHS algorithm and `classic' SPH. We describe our initial conditions, and present our implementation of radiative cooling, star formation and stellar feedback, and our treatment of a central supermassive black hole (SMBH). In Sections \ref{sec:results} \& \ref{sec:fullpower} we present our results, which we discuss in Section \ref{sec:discussion}. Finally, in Section \ref{sec:conclusions}, we present our conclusions.

\section{Methods}\label{sec:method}

\subsection{SPHS}\label{sec:SPHS} 

The SPHS method is described in detail in \cite{2009arXiv0906.0774R} and \cite{2012MNRAS.tmp.2941R}. We give a brief summary of the main equations here. The discretised hydrodynamic equations of motion are chosen to minimise force errors as in \cite{2009arXiv0906.0774R}: 

\begin{equation}
\rho_i = \sum_j^N m_j \wij(|{\bf r}_{ij}|,h_i)
\label{eqn:sphcont}
\end{equation}
\begin{equation}
\frac{d\uv_i}{dt} = -\sum_j^N \frac{m_j}{\rho_i\rho_j} \left[P_i + P_j\right] \nabla_i \overline{W}_{ij}
\label{eqn:sphmoment}
\end{equation}
\begin{equation}
P_i = A_i \rho_i^\gamma
\label{eqn:sphstate}
\end{equation}
where $m_i$ is the mass of particle $i$; $\overline{W}_{ij} = \frac{1}{2}\left[W_{ij}(h_i) + W_{ij}(h_j)\right]$; and $W$ is a symmetric kernel that obeys the normalisation condition:
\begin{equation}
\int_{V} W(|\ur-\ur'|,h) \dthr' = 1
\label{eqn:normw}
\end{equation}
and the property (for smoothing length $h$):
\begin{equation}
\lim_{h\rightarrow 0} W(|\ur-\ur'|,h) = \delta(|\ur-\ur'|)
\end{equation}
and ${\bf r}_{ij} = {\bf r}_j - {\bf r}_i$ is the vector position of the particle relative to the centre of the kernel.

We use a variable smoothing length $h_i$ as in \cite{SpringelHernquist02} that is adjusted to obey the following constraint equation: 

\begin{equation}
\frac{4\pi}{3} h_i^3 n_i = N_n \qquad ;\,\mathrm{with} \qquad n_i = \sum_j^N W_{ij}
\label{eqn:fixedmass}
\end{equation}
where $N_n$ is the typical neighbour number (the number of particles inside the smoothing kernel, $W$). The above constraint equation gives fixed mass inside the kernel if particle masses are all equal. For the full SPHS runs (see Section \ref{sec:fullpower}) we use the `HOCT4' kernel with 442 neighbours as this gives significantly improved force accuracy and convergence \citep{2009arXiv0906.0774R,2012MNRAS.tmp.2941R}:

\begin{equation}
W = \frac{N}{h^3}\left\{\begin{array}{lr}
Px + Q & 0 < x \le \kappa \\
(1-x)^4 + (\alpha - x)^4 + (\beta-x)^4 & \kappa < x \le \beta \\
(1-x)^4 + (\alpha - x)^4 & \beta < x \le \alpha\\
(1-x)^4 & x \le 1 \\
0 & \mathrm{otherwise} \end{array}\right.
\label{eqn:hoctkern}
\end{equation}
with $N = 6.515$, $P=-2.15$, $Q=0.981$, $\alpha = 0.75$, $\beta = 0.5$ and $\kappa = 0.214$.

In addition to the above equations of motion, numerical dissipation is switched on if particles are converging. This avoids multivalued fluid quantities occurring at the point of convergent flow. Without such dissipation, the resulting multivalued pressures drive waves through the fluid that propagate large numerical errors and spoil convergence. The switch is given by: 

\begin{equation}
\alpha_{\mathrm{loc},i} = \left\{
\begin{array}{lr}
\frac{h_i^2 |\unabla(\unabla \cdot {\bf v}_i)|}{h_i^2 |\unabla(\unabla \cdot {\bf v}_i)| + h_i |\unabla \cdot {\bf v}_i|+ n_s c_i} \alpha_\mathrm{max} & \unabla \cdot {\bf v}_i < 0 \\
0 & \mathrm{otherwise} 
\end{array}\right.
\label{eqn:alphalocv}
\end{equation}
where $\alpha_{\mathrm{loc},i}$ describes the amount of dissipation for a given particle in the range $[0,\alpha_\mathrm{max} = 1]$; $c_i$ is the sound speed of particle $i$; and $n_s = 0.05$ is a `noise' parameter that determines the magnitude of velocity fluctuations that trigger the switch. Equation \ref{eqn:alphalocv} turns on dissipation if $\unabla \cdot {\bf v}_i < 0$ (convergent flow) and if the magnitude of the spatial derivative of $\unabla \cdot {\bf v}_i$ is large as compared to the local divergence (i.e., if the flow is going to converge). The key advantage as compared to most other switches in the literature is that it acts as an early warning system, switching on {\it before} large numerical errors propagate throughout the fluid \citep[see also][]{2010MNRAS.408..669C}. The second derivatives of the velocity field are calculated using high order polynomial gradient estimators described in \cite{2003ApJ...595..564M} and \cite{2012MNRAS.tmp.2941R}. We use the above switch to turn on dissipation in all advected fluid quantities -- i.e., the momentum (artificial viscosity) and entropy (artificial thermal conductivity). Once the trajectories are no longer converging, the the dissipation parameter decays back to zero on a timescale $\sim h_i / c_i$. The dissipation equations are fully conservative and described in detail in \cite{2012MNRAS.tmp.2941R}. The only free parameter is $\alpha_\mathrm{max}$ that describes the rate of dissipation that occurs when particle trajectories attempt to cross. Note that by construction, the numerical dissipation occurs at the resolution limit of the simulation and so convergence is independent of $\alpha_\mathrm{max}$.

In addition to the hydrodynamical modifications, the SPHS method also includes an improved timestepping algorithm for strong shocks. This was adapted from \cite{SaitohMakino2009}, who found that the evolution of a Sedov-Taylor blast wave (or similar) is captured incorrectly when the gas in the shock is on a very different timestep to the gas it is impinging upon\footnote{We would like to thank Frazer Pearce \& Stewart Muldrew for providing the first version of the timestepping code patch that we use.}. In extreme cases, the particles sitting ahead of the shock may not `feel' the particles in the blast wave, as the stationary particles are on such long timesteps the shock has passed by before their next force update. \cite{SaitohMakino2009} recommend restricting the timesteps between neighbours to be at most a factor of 4 in order to alleviate this problem; this is what we do also in SPHS.

The SPHS algorithm was incorporated into the N-body/hydrodynamical code {\tt GADGET-3} \citep{Springel2005}, which was used for all the simulations in this paper.

\subsection{Classic SPH}\label{sec:classic} 

Throughout this paper, we will present comparisons with `classic' SPH. We define this to be the fully conservative `entropy' form of SPH described in \cite{SpringelHernquist2002}. However, many production SPH codes -- e.g. {\tt Gasoline} \citep{2004NewA....9..137W}, and {\tt Hydra} \citep{1995ApJ...452..797C} -- are sufficiently similar to warrant also being described as `classic' SPH. The discretised Euler equations are the same as in SPHS, but with the momentum equation replaced by:  

\begin{equation}
\frac{d\uv_i}{dt} = -\sum_j^N m_j \left[f_i \frac{P_i}{\rho_i^2} \nabla_i W_{ij}(h_i) + f_j \frac{P_j}{\rho_j^2} \unabla_i W_{ij}(h_j)\right]
\label{eqn:sphmomentcons}
\end{equation}
where the function $f_i$ is a correction factor that ensures energy conservation for varying smoothing lengths:
\begin{equation}
f_i = \left(1 + \frac{h_i}{3\rho_i}\frac{\partial \rho_i}{\partial h_i}\right)^{-1};
\end{equation}
We do not use the above conservative momentum equation in SPHS since it leads to larger force errors with only a modest improvement in energy conservation (at least when applied to galaxy and galaxy cluster formation simulations; see \cite{2009arXiv0906.0774R} and \cite{2012MNRAS.tmp.2941R} for further details).

Unlike in SPHS, there is no dissipation switching and $\alpha = \alpha_\mathrm{max} = \mathrm{const.} = 1$ always. There is also no dissipation in entropy; the only numerical dissipation applied is the artificial viscosity. This prevents multivalued momenta from occurring, but not multivalued entropy or pressure \citep[e.g.,][]{2009arXiv0906.0774R}.

For most simulations presented in this paper, we use a standard cubic spline (CS) kernel with 96 neighbours -- both for SPH and SPHS. This is somewhat larger than usually employed but is necessary for the high order gradients required for the dissipation switch in SPHS \citep{2012MNRAS.tmp.2941R}. By using the same neighbour number for both hydrodynamic techniques, we ensure like spatial resolution. However, at this neighbour number, the CS kernel is prone to a pairing instability \citep{ReadEtal2010} that could in principle seed numerical instabilities in a hot corona. We explicitly check that this is not the case by running one of our classic SPH simulations with a more standard 42 neighbours, and with the HOCT4 kernel with 442 neighbours (that is manifestly stable to pairing). In all cases, we recover the result seen in the literature that the hot coronae breaks up into many cold clumps.

\subsection{Initial conditions}\label{sec:ic}

Our setup is similar to the `cooling gaseous halo' model of \cite{KaufmannEtal2006} and \cite{KaufmannEtal2007}. For our initial condition, we employ a live dark matter (DM) halo of collisionless particles together with a gaseous halo of SPH particles, with each component relaxed for many dynamical times with an adiabatic equation of state (EQS) in order to remove Poisson noise. The relaxation step is performed separately for each resolution and kernel-neighbour number combination used. Both the DM and gas distributions follow a Dehnen-McLaughlin model \citep{DehnenMcLaughlin2005} of the form
\begin{equation}
\rho(r) \propto \frac{1}{(r/r_s^{7/9}) \left[1 + (r/r_s^{4/9}) \right]^6}
\label{eq:rho}
\end{equation}
with the gas profile normalised to contain a mass of $1.5 \times 10^{11} \msun$ and the DM to contain a mass of $1.5 \times 10^{12} \msun$. The scale radius for the haloes is $r_s = 40$ kpc, and both are truncated at a virial radius $r_t = 200$ kpc.  

The gas is initially (after relaxation) in hydrostatic equilibrium, with a temperature profile given by the relation
\begin{equation}
T(r) = \frac{\mu m_p}{k_{\rm B}}\frac{1}{\rho_{\rm gas} (r)} \int_r^\infty \rho_{\rm gas} (r) \frac{GM(r)}{r^2} \, dr
\end{equation}
where $\mu$ is the mean molecular weight, $k_{\rm B}$ is the Boltzmann constant, $\rho_{\rm gas} (r)$ is the radial gas density profile and $M(r)$ is the enclosed mass of both components within a radius $r$. The gas is given a velocity field whereby the specific angular momentum profile follows a power-law \citep{BullockEtal2001b, KaufmannEtal2007} such that
\begin{equation}
j_{\rm gas} \propto r^{1.0}
\end{equation}
and normalised by a rotation parameter $\lambda = 0.038$, defined by
\begin{equation}
\lambda = \frac{j_{\rm gas} \vert E_{\rm dm} \vert^{1/2}}{GM_{\rm dm}^{3/2}}
\end{equation}
where $E_{\rm dm}$ and $M_{\rm dm}$ are the total energy and mass of the DM halo. This normalisation implicitly assumes negligible angular momentum transport between the DM halo and the gas, an assumption justified in \cite{KaufmannEtal2007}. The rotational velocity field is implemented about the $z$-axis.

Relevant parameters for the simulations are given in Table \ref{tableics}. The DM haloes were constructed using variable particle masses in order to minimise computational expense, as described in Section 3.1 of \cite{ColeEtal2011}. Gravitational softening lengths for the collisionless N-body components (DM \& stars) were fixed, with the values given in Table \ref{tableics} ($\epsilon_{\rm dm}$, which also applies to the softening lengths of any stars formed during the simulation). For the gas, we employed adaptive gravitational softening lengths which were set equal to the gas smoothing length at all times so as to ensure there was no artificial bias towards pressure support or gravitational collapse in Jeans-unstable regions, as per \cite{BateBurkert1997}. We did not employ the conservative correction terms as suggested in \cite{Price2007}; we will explore these in future work. To explicitly check that our variable softening does not affect our results, we re-ran the SPH-96-res1 and SPHS-96-res1 simulations with fixed gravitational softening in the gas, set to the minimum smoothing length in the latter two simulations. Our results were largely unchanged in terms of the SPH/SPHS comparison.
 
\begin{table*}
\caption{Parameters for each simulation. Two resolutions were run for each numerical method. $N$ is the total number of particles for each species, $m_{\rm gas}$ is the (constant) mass of each gas particle and $m_{\rm dm}$ gives the range of masses for the dark matter particles. $M_{\rm res}$ is the `resolvable' gas mass, defined in Section \ref{sec:cooling}. $h$ refers to the smoothing length of the gas while $\epsilon$ refers to the gravitational softening length. For all simulations, $\epsilon = h$ for the gas particles.}
\centering
\begin{tabular}{|c|c|c|c|c|c|c|c|c|}\hline
ID & $N_{\rm gas}$ & $N_{\rm dm}$ & $m_{\rm gas}$ ($\msun$) & $m_{\rm dm}$ ($\msun$) & $M_{\rm res}$ ($\msun$) & $h_{\rm min}$ (kpc) & $h_{\rm max}$ (kpc) & $\epsilon_{\rm dm}$ (kpc) \\ \hline
\hline
SPH-96-res0 & $1.5 \times 10^5$ & $1.8 \times 10^5$ & $1 \times 10^6$ & $1 - 14 \times 10^6$  & $1.28 \times 10^8$ & $0.1$ & $1790$ & $0.45$ \\
SPHS-96-res0 & $1.5 \times 10^5$ & $1.8 \times 10^5$ & $1 \times 10^6$ & $1 - 14 \times 10^6$ & $1.28 \times 10^8$ & $0.08$ & $295$ & $0.45$ \\
SPH-96-res1 & $7.5 \times 10^5$ & $9 \times 10^5$ & $2 \times 10^5$ & $2 - 29 \times 10^5$ & $2.56 \times 10^7$ & $0.03$ & $310$ & $0.2$ \\
SPHS-96-res1 & $7.5 \times 10^5$ & $9 \times 10^5$ & $2 \times 10^5$ & $2 - 29 \times 10^5$ & $2.56 \times 10^7$ & $0.03$ & $230$ & $0.2$ \\
\hline
SPHS-442-res2 & $3.75 \times 10^6$ & $4.5 \times 10^6$ & $4 \times 10^4$ & $4 - 59 \times 10^4$ & $5.12 \times 10^6$ & $0.008 $ & $200$ & $0.03$ \\
\hline
\hline
\end{tabular}
\begin{flushleft}
\end{flushleft}
\label{tableics}
\end{table*}

\subsection{Radiative cooling}\label{sec:cooling}

For all of the simulations, we employ radiative cooling in the gas with a cooling floor of $T_{\rm floor} = 100$\,K. We use a toy cooling curve that follows \cite{KWH1996} above $10^4$\,K, assuming primordial abundance, and \cite{MashchenkoEtal2008} below $10^4$\,K assuming Solar abundance. This crudely models a low metallicity cooling halo that rapidly reaches $\sim$ Solar abundance in cooling star forming regions. We will consider a more realistic cooling curve that self-consistently responds to the injection of metals from star forming regions in future work.

In addition to a cooling floor, we also prevent gas from cooling to the point at which the Jeans mass for gravitational collapse becomes unresolved. A given SPH/SPHS simulation has a mass resolution equal to $M_{\rm res} = N_{\rm res} m_{\rm gas}$, where $m_{\rm gas}$ is the mass of a gas particle and $N_{\rm res}$ is the number of gas particles that constitutes a resovable mass, i.e., a single resolution element. \cite{BateBurkert1997} find this to be approximately $2 N_{\rm neigh}$, where $N_{\rm neigh}$ is the number of neighbours of a gas particle; this number is somewhat of a rule of thumb and depends on the resolving scale of the particular kernel employed. For all of the tests we use $N_{\rm res} = 128$. The resolving scales of our CS and HOCT4 kernels (see Section \ref{sec:SPHS}) are such that this number is reasonable \citep[for more detail on the resolving power of these kernels see][]{ReadEtal2010, 2012MNRAS.tmp.2941R, DehnenAly2012}. 

Our `dynamic' cooling floor effectively ensures that the Jeans mass is always resolved within our simulation. For a given mass element $M_{\rm res}$, we can write a Jeans density, namely,
\begin{equation}
\rho_J = \left(\frac{\pi k T}{\mu m_p G}\right)^3 \left(\frac{1}{M_{\rm res}}\right)^2
\label{eq:polytrope}
\end{equation}
which manifests in the simulation as a polytropic equation of state $P = A(s) \rho^{4/3}$. Gas is not allowed to collapse (for a given temperature) to densities higher than given by equation \ref{eq:polytrope}, and we identify gas that lies on the polytrope as `star-forming', allowing it to form stars above a fixed density threshold (see Section \ref{sec:sf}). We emphasise that the polytrope is not physical, but a purely numerical device to prevent star-forming gas from becoming unresolved in our simulations \citep[see, e.g.,][]{TrueloveEtal1997, vandeVoortEtal2011, DuboisEtal2012}.

\subsection{Star formation and feedback}\label{sec:sf}

Star formation (SF) in our simulations is modelled in a `sub-grid' fashion according to observational constraints on the SF rate and efficiency. We allow gas that lies on the polytrope to form stars (N-body particles) above a fixed density threshold, with an efficiency of 0.1 as per observations \citep[e.g.,][]{LadaLada2003} of giant molecular clouds (GMCs). The star formation rate follows the Schmidt volume density law for star formation \citep{Schmidt1959}, namely
\begin{equation}
\rho_{\rm SFR} \propto \rho_{\rm gas}^{1.5}
\end{equation}
which is implemented in the simulation by employing the dynamical time as the relevant SF timescale, i.e.
\begin{equation}
\frac{\text{d} \rho_*}{\text{d}t} = \eta \frac{\rho_{\rm gas}}{t_{\rm dyn}}
\end{equation}
where $\eta$ is the SF efficiency.

In all the simulations, we include feedback from supernovae (SNe), which takes the form of an injection of thermal energy from the star particle into nearby gas particles. Due to our finite resolution, each star particle is not an individual star but is instead representative of a stellar distribution, and so we integrate over a Salpeter IMF \citep{Salpeter1955} between $8 \msun$ and $100 \msun$ in order to determine the number of SNe that should be feeding back at any particular time. We treat only type II SNe, with the energy of each supernova event set to $E_{\rm SN} = 10^{51}$ ergs. To determine the time of injection we use the standard relation:
\begin{equation}
\frac{t_{\rm MS}}{\tsun} \sim \left(\frac{M}{\msun}\right)^{-2.5}
\end{equation}
to determine an approximate main sequence time $t_{\rm MS}$ after the initial formation, at which the energy injection is implemented as a delta function in time. We couple the feedback energy to a given mass of gas, rather than just the neighbours of the star particle -- this mass is set to be our resolvable mass $M_{\rm res}$ (see Section \ref{sec:cooling}). In doing so, we ensure that both the star formation and the feedback is resolved. The actual injection of energy is convolved with a CS kernel to ensure a smoothing of the thermal coupling over the mass scale receiving the supernova feedback.

Th polytropic pressure floor, together with the fixed SF density threshold, is crucial to our sub-grid modelling strategy for star formation; it means that the gas that is able to form stars is (i) undergoing Jeans collapse and (ii) at the typical density of GMC star-forming regions. By constraining `star-forming gas' to lie on the polytrope we are ensuring that star formation occurs in collapsing gas regions, rather than just in regions of high density. We note that there has been some recent evidence that GMCs and GMC complexes may not need to be globally gravitationally bound for regions within them to undergo star formation \citep{DobbsEtal2011} but that even in this picture the dense sub-structures actually forming stars are both (i) in a state of gravitational collapse and (ii) at a density of $\simgt 10^2$ atoms cm$^{-3}$. 

A further aspect of our polytropic pressure floor is to maintain consistency with the parameters of the SF sub-grid model. These are tailored to star-forming gas at approximately a GMC mass scale, and employ the corresponding global observational constraints, rather than including any of the finer details of the star formation process (e.g., metal-line cooling, non-equilibrium chemistry, etc.). To allow gas to collapse to higher and higher densities as it cools would lead to a situation whereby not only would this collapse become dominated by numerical effects but also reach a state that was inconsistent with the global constraints of our sub-grid prescription. We plan to explore our sub-grid modelling strategy in this context in a forthcoming paper (Hobbs et al., in preparation), with particular importance placed on convergence with increasing resolution. In the current paper we do not expect convergence of the star formation properties owing to the fact that the polytrope line shifts to lower mass scales as the resolution is increased; this however is not a problem for the comparison we make between the two numerical methods, since this is done at identical resolution. We note that such a `non-convergent' approach is similar to what is typically used in other galaxy formation sub-grid models in the literature.

\subsection{SMBH sub-grid model}

We model the SMBH at the centre of the halo as a sink particle \citep[see, e.g.,][]{Bate95}. The accretion of gas onto the SMBH is implemented using an `accretion radius' approach; gas that makes it inside a distance of $r_{\rm acc} = 0.1$ kpc from the black hole is removed from the simulation, while its mass is then added to the SMBH mass $M_{\rm bh}$. We start the SMBH as a seed of negligible mass since even the accretion of a single gas particle grows it to a mass of $\sim 10^4 - 10^6 \msun$ depending on the resolution (see Table 1). As a result of this mass being negligible before any accretion occurs, it was necessary to tie the black hole to the centre of the potential in order to stop it wandering excessive distances during the early phases of accretion. We therefore re-position the SMBH on the centre of the potential at every timestep.

For the simulations presented in this paper we do not model feedback from the SMBH, using it as a sink for accreting gas only. Our prescription for SMBH growth in the present paper is somewhat crude, but we stress that this is because our focus is on the differences between the SPH and SPHS methods in terms of thermal instabilities in the hot halo gas; keeping our SMBH sub-grid model simple allows us to present a cleaner numerical test between the two hydrodynamic methods. We plan to investigate the role of black hole feeding and feedback in similar galaxy formation simulations with a more advanced sub-grid model in future papers.

\section{Results}\label{sec:results}

We run four main simulations, two in classic SPH, and two in SPHS each at two different resolutions (see Table \ref{tableics}). The global properties between the different runs are similar due to the identical initial conditions; as the gaseous halo cools it loses hydrostatic support and begins to collapse, with the small net rotation velocity about the $z$-axis causing a disc to form initially in the $x-y$ plane. We start by describing the overall behaviour of the simulations, and then discuss the differences in the results between the two numerical methods and any resolution-dependant departures from this overall behaviour. Our fiducial comparison is made between the two higher resolution runs, SPH-96-res1 and SPHS-96-res1.

As the gaseous halo collapses, it falls towards the centre of the computational domain. Within the first $\sim 10$ million years, some of the gas that has reached the central $\sim 1$ kpc undergoes a starburst, while the rest is accreted by the SMBH. The feedback associated with this star formation event is strong enough to drive a near-spherical overpressurised bubble outwards into the infalling gas, temporarily evacuating the central region of most of its gas and halting the infall within the expanding cavity that is created. Star formation largely ceases at this point, while the gas that has formed into a small-scale disc in the central kiloparsec continues to accrete at a significantly lower rate. The expanding shell sweeps up mass and slows down, eventually allowing further infall to the central regions, and in most cases, subsequent starbursts and feedback events.

While the initial starburst was largely spherical, driving a very symmetric bubble into the surrounding gas, the subsequent feedback events occur with a greater degree of asymmetry, and so hot bubbles are generated at a variety of different orientations at different times. In these later events gas is able to flow in more readily along the intersections between bubbles, where it is able to build up the disc in the central few kiloparsecs.

\subsection{Differences between SPH and SPHS}\label{sec:differences}

\subsubsection{Lower resolution (res0)}

\begin{figure*}
\begin{minipage}[b]{.49\textwidth}
\centerline{\psfig{file=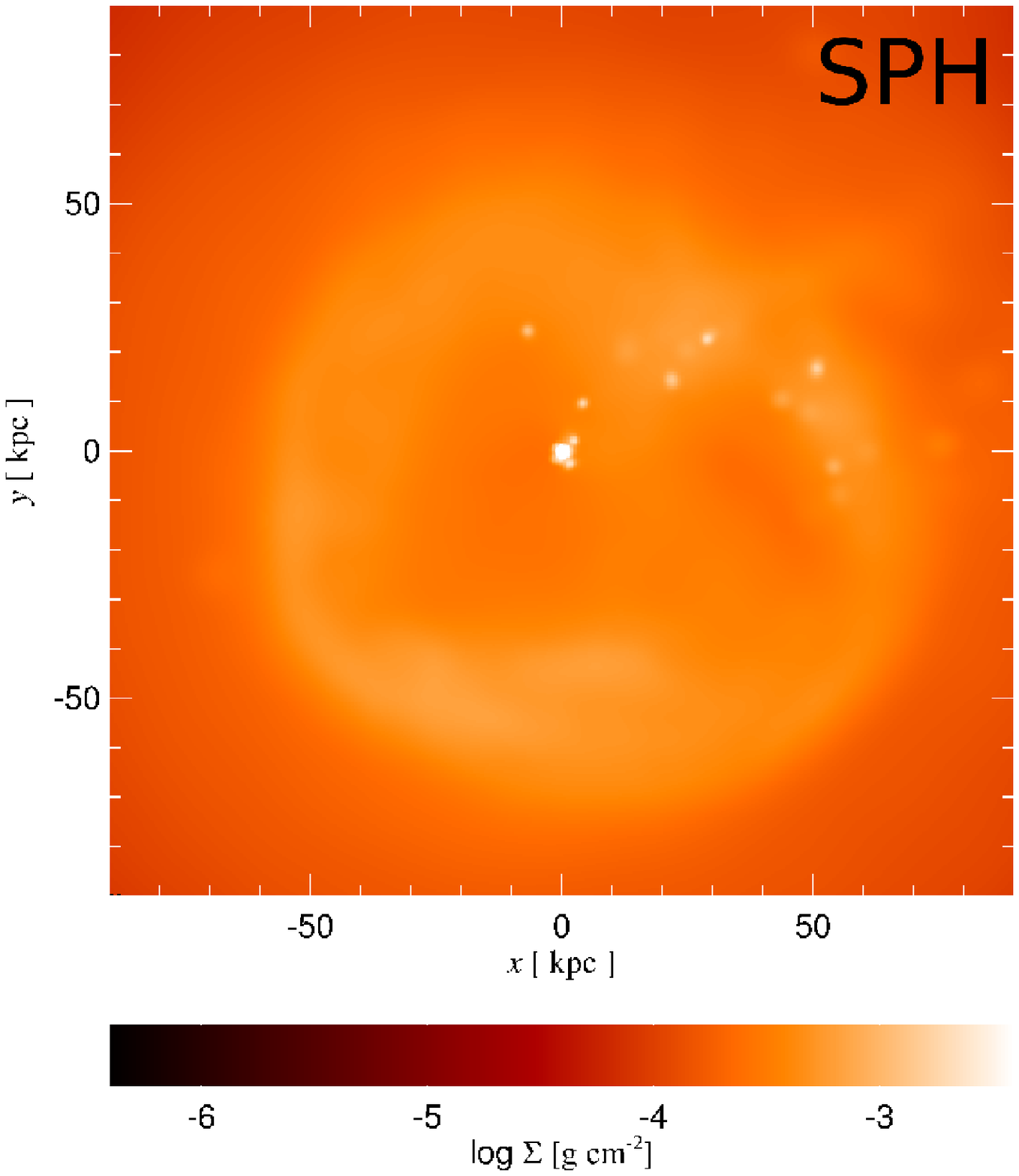,width=1.00\textwidth,angle=0}}
\end{minipage}
\begin{minipage}[b]{.49\textwidth}
\centerline{\psfig{file=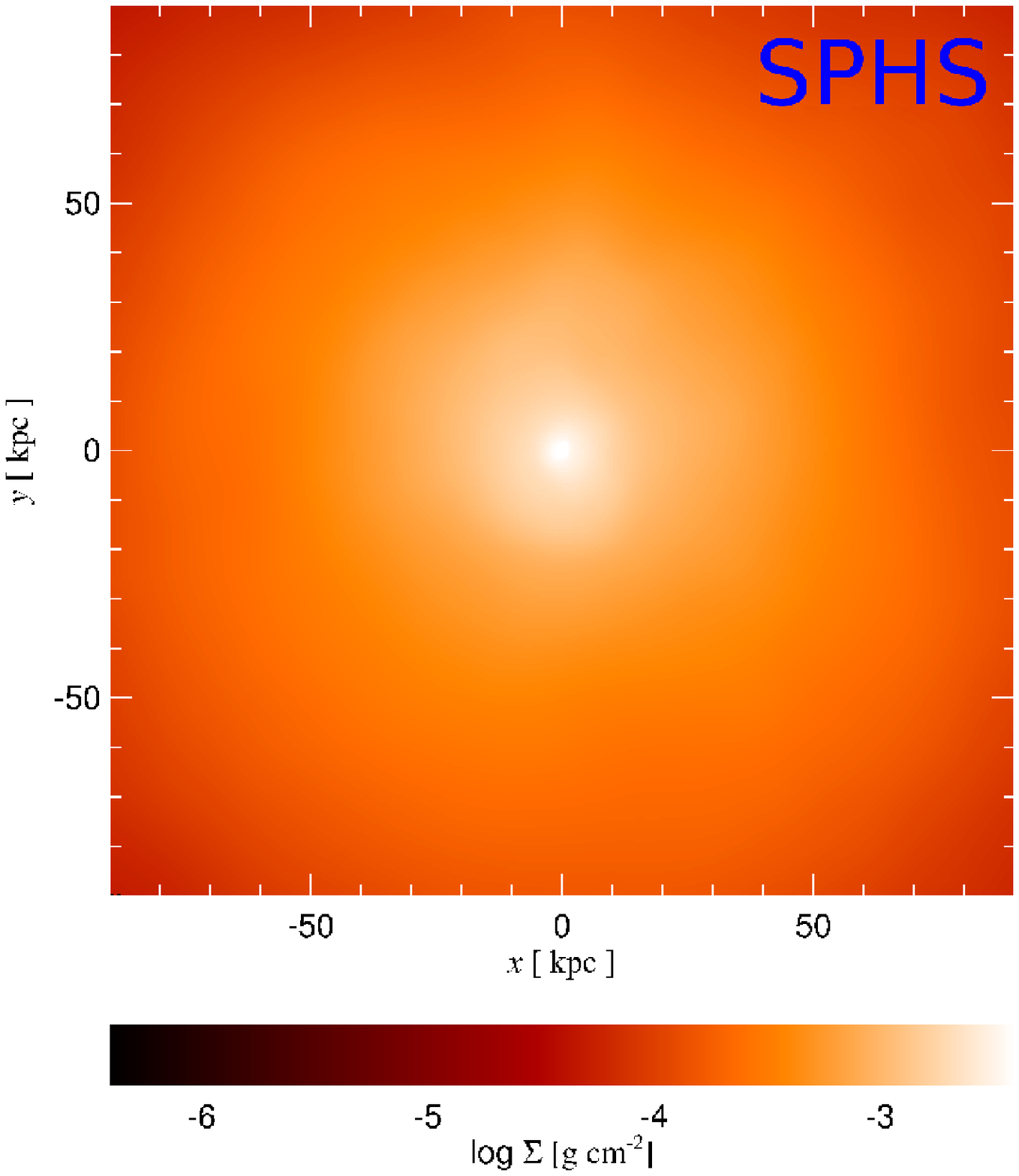,width=1.00\textwidth,angle=0}}
\end{minipage}
\caption{Projected surface density plots for SPH-96-res0 (left) and SPHS-96-res0 (right), both with $N_{\rm gas} = 1.5 \times 10^5$, at $t = 2$\,Gyr. There is a clear difference in the amount of structures present in the hot halo due to supernovae feedback - in SPH the shape of the overpressurised bubbles is more defined and a number of spherical clumps have condensed out of the ambient gas. In SPHS the gas distribution is more homogeneous, and no such clumps have formed.}
\label{fig:snapshotinfall2e5}
\end{figure*}

\begin{figure*}
\begin{minipage}[b]{.49\textwidth}
\centerline{\psfig{file=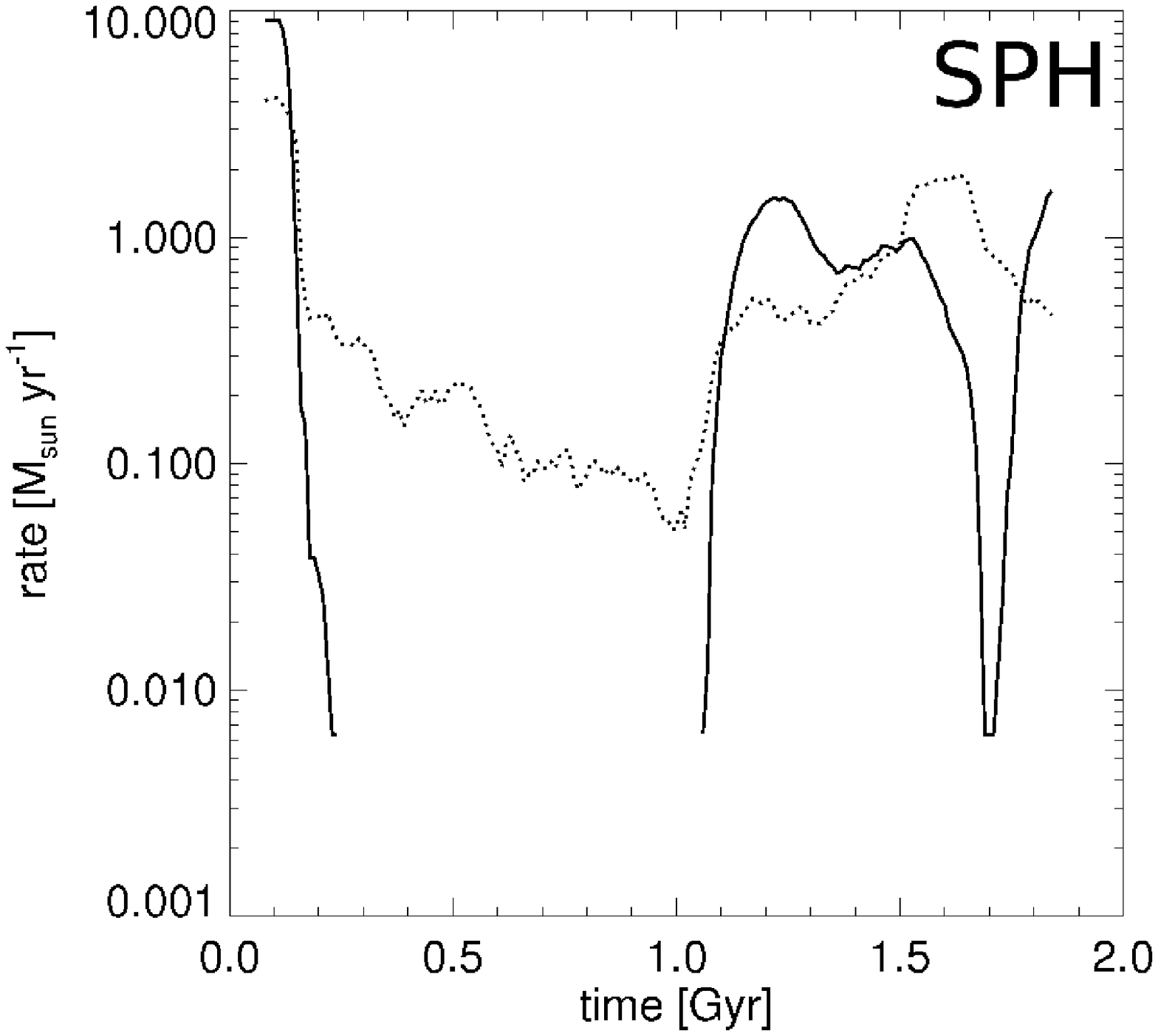,width=1.00\textwidth,angle=0}}
\end{minipage}
\begin{minipage}[b]{.49\textwidth}
\centerline{\psfig{file=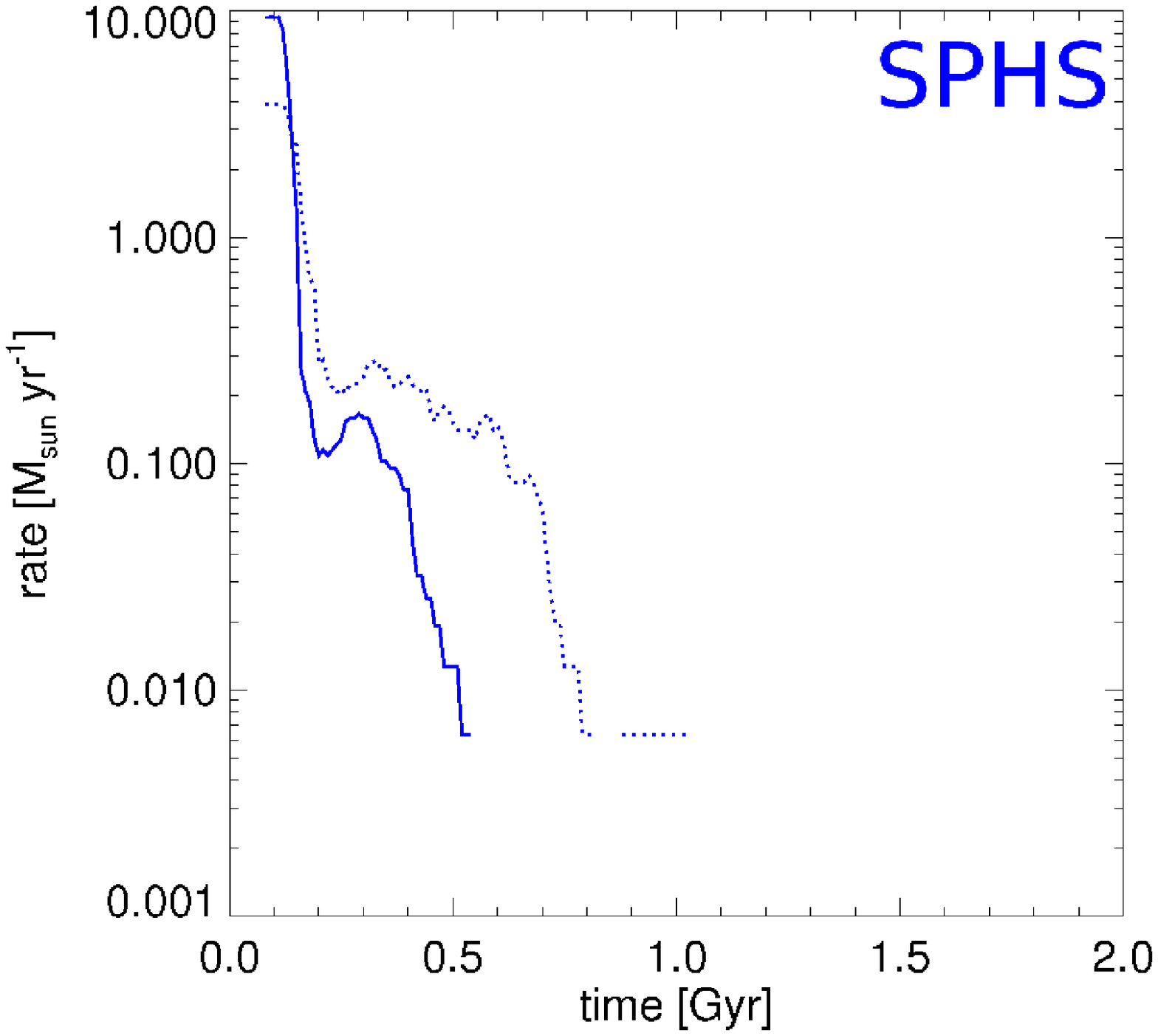,width=1.00\textwidth,angle=0}}
\end{minipage}
\caption{SFR and SMBH accretion histories for $N_{\rm gas} = 2 \times 10^5$, with SPH (left) and SPHS (right). Solid line is SFR, dotted is $\dot{M}_{\rm bh}$. The initial starbursts and accretion events start off identical, but in SPHS both the SFR and $\dot{M}_{\rm bh}$ are prolonged compared to SPH, with a secondary peak just as the main starburst is shutting off. The SPH run shows a second major starburst not present in the SPHS run. In both simulations the shape of the SMBH accretion rate curve largely follows that of the SFR, albeit with a small offset in time.}
\label{fig:sfrsmbh2e5}
\end{figure*}

We start by comparing the low resolution runs, SPH-96-res0 and SPHS-96-res0. These are shown at a time of $t = 2$\,Gyr in Figure \ref{fig:snapshotinfall2e5}. The initial evolution is similar, with some of the gas that reaches the central kiloparsec being accreted by the SMBH, and some undergoing a starburst that drives a hot bubble into the gaseous halo. In the SPHS case, this initial feedback is slightly stronger, with the gas that has received the thermal `kick' reaching higher temperatures - a maximum of $5 \times 10^8$ K vs. $8 \times 10^7$ K - and expanding slightly faster than in the SPH case. Immediately after the feedback event, the accretion rate drops by several orders of magnitude (see Figure \ref{fig:sfrsmbh2e5}). The expanding shell slows down as it sweeps up mass from the gaseous halo, and it is at this point that the two simulations differ; in the SPH run the gas is able to make its way back inside the central few kiloparsecs, where it cools and forms stars, giving rise to a second starburst at $\sim 1.2$\,Gyr. In the SPHS case, however, the gas has not collapsed back into the central regions by the end of the simulation at $t = 2$\,Gyr, and so there is no associated subsequent starburst event by this time.

We can understand some of these differences by looking at the star formation history between the two low resolution runs, which can be seen in Figure \ref{fig:sfrsmbh2e5} along with the accretion history. The SMBH accretion rate largely follows the star formation rate, and so in the SPH run there are 2-3 major starburst and accretion events that take place to later times after the initial one. In the SPHS case the initial starburst lasts for longer -- rather than undergoing a sudden event and then dropping to zero, as it does in SPH, the star formation rate shows a more sustained period of activity for the first $\sim 0.5$\,Gyr. This is the reason for the lack of the second starburst in the SPHS run - the sustained initial star formation and feedback has driven the gas to higher temperatures and lower densities and so it takes a much longer time to cool and condense back into the central regions, failing to do so by the end of the simulation at $t = 2$\,Gyr.

Crucially, even at this low resolution we can see that by far the main difference between the SPH and SPHS simulations is the presence of overdense clumps that have condensed from the ambient halo gas between $\sim 10 - 50$\,kpc. These clumps are not present in the SPHS run, and we go into more detail on this particular result for the higher resolution case in Section \ref{sec:clumps}.

\subsubsection{Higher resolution (res1)}

\begin{figure*}
\begin{minipage}[b]{.49\textwidth}
\centerline{\psfig{file=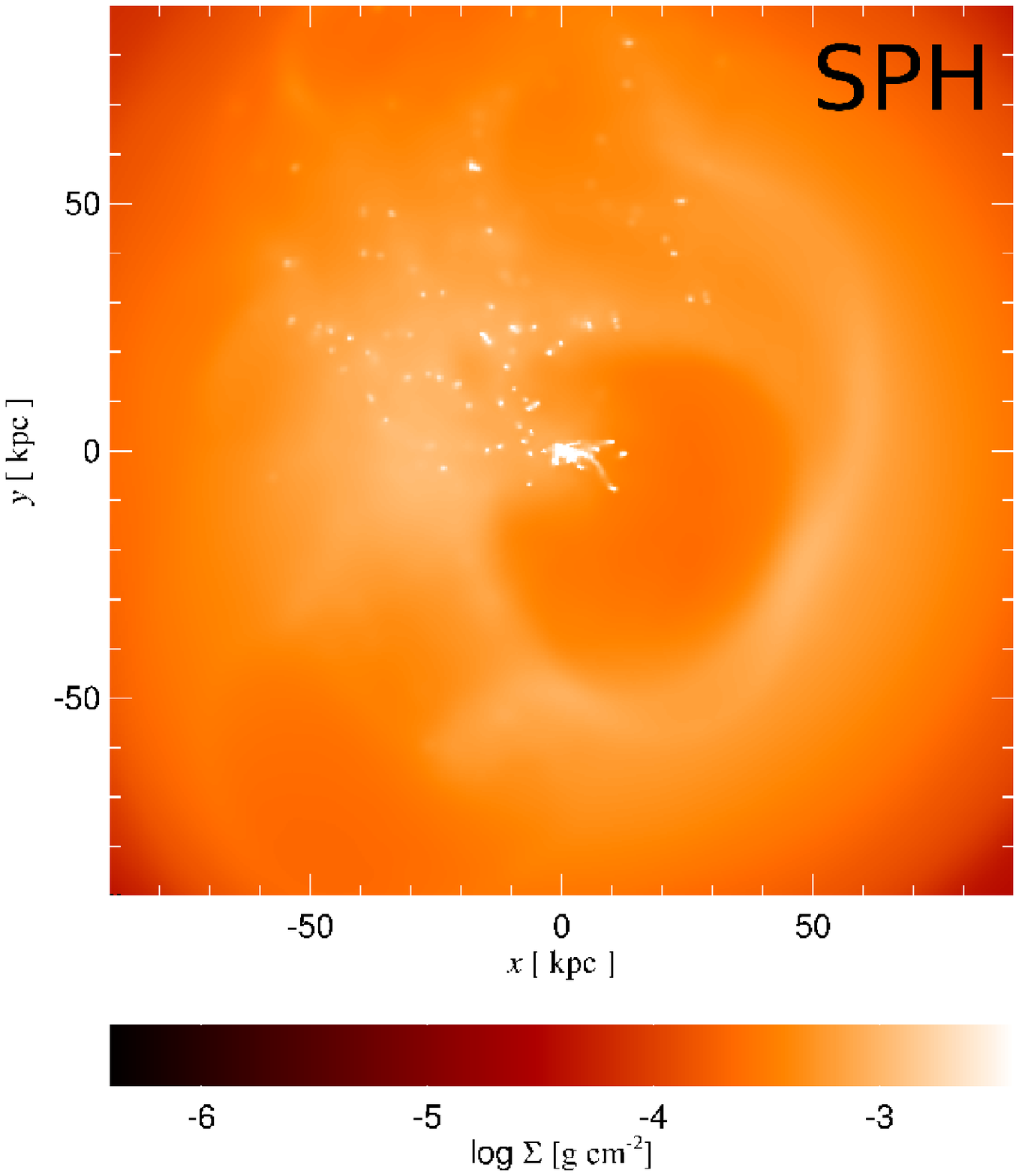,width=1.00\textwidth,angle=0}}
\end{minipage}
\begin{minipage}[b]{.49\textwidth}
\centerline{\psfig{file=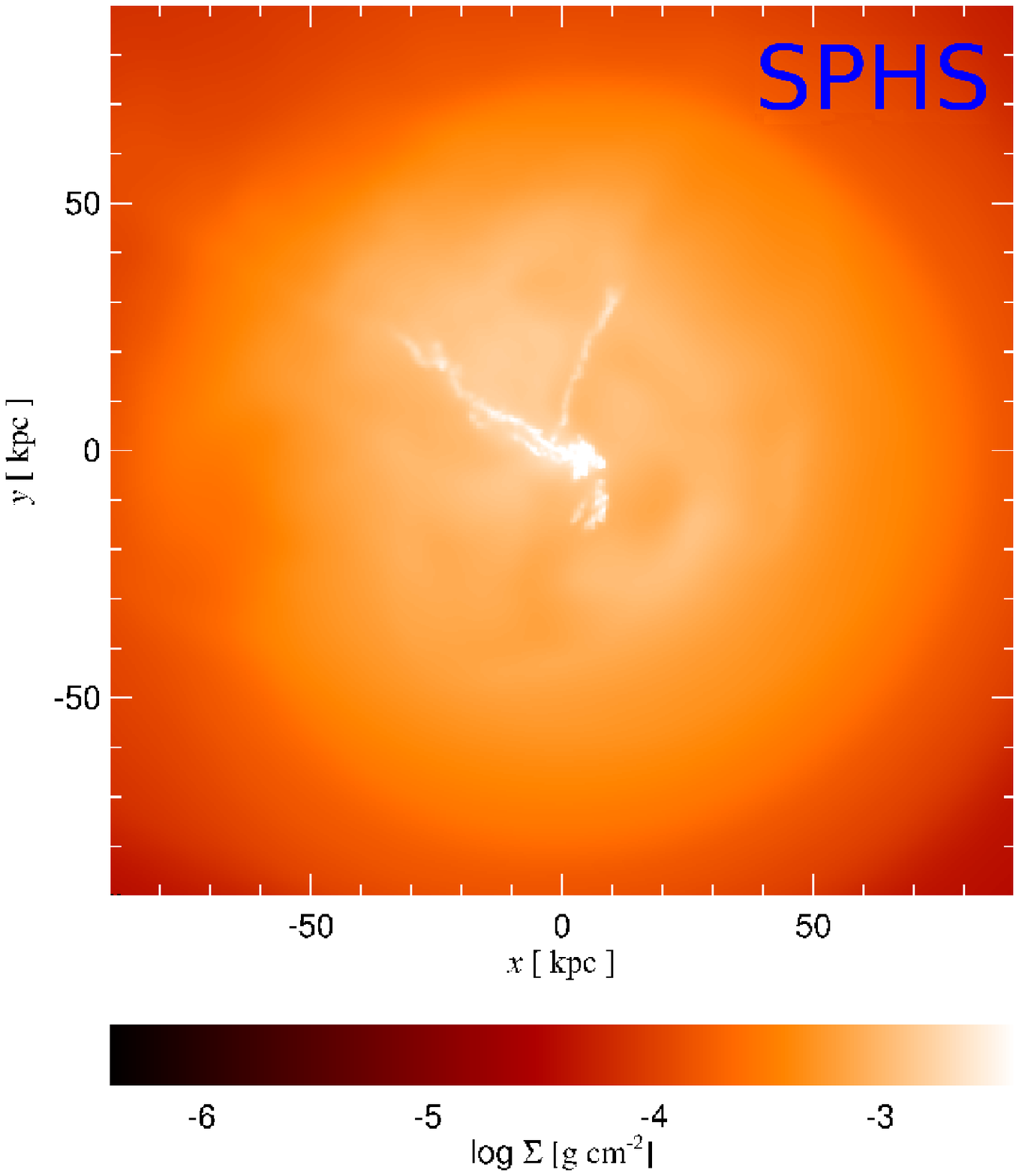,width=1.00\textwidth,angle=0}}
\end{minipage}
\caption{Projected surface density plots for SPH-96-res1 (left) and SPHS-96-res1 (right), both with $N_{\rm gas} = 7.5 \times 10^5$, at $t = 1.2$\,Gyr. While there is structure in both runs at this resolution, the mode of structure formation is very different, with the gas that has cooled to $\sim 10^4$\,K condensing into a myriad of clumps in the SPH run but forming into two distinct filaments in the SPHS run. As a result, the manner in which the central gaseous disc grows is different; in SPH some of the clumps scatter off one another and fall into the centre to feed the disc, whereas in SPHS the disc grows as gas is funnelled down through the dense filaments that form due to the intersection of two or more supernovae-driven bubbles.}
\label{fig:snapshotinfall1e6}
\end{figure*}

\begin{figure*}
\begin{minipage}[b]{.49\textwidth}
\centerline{\psfig{file=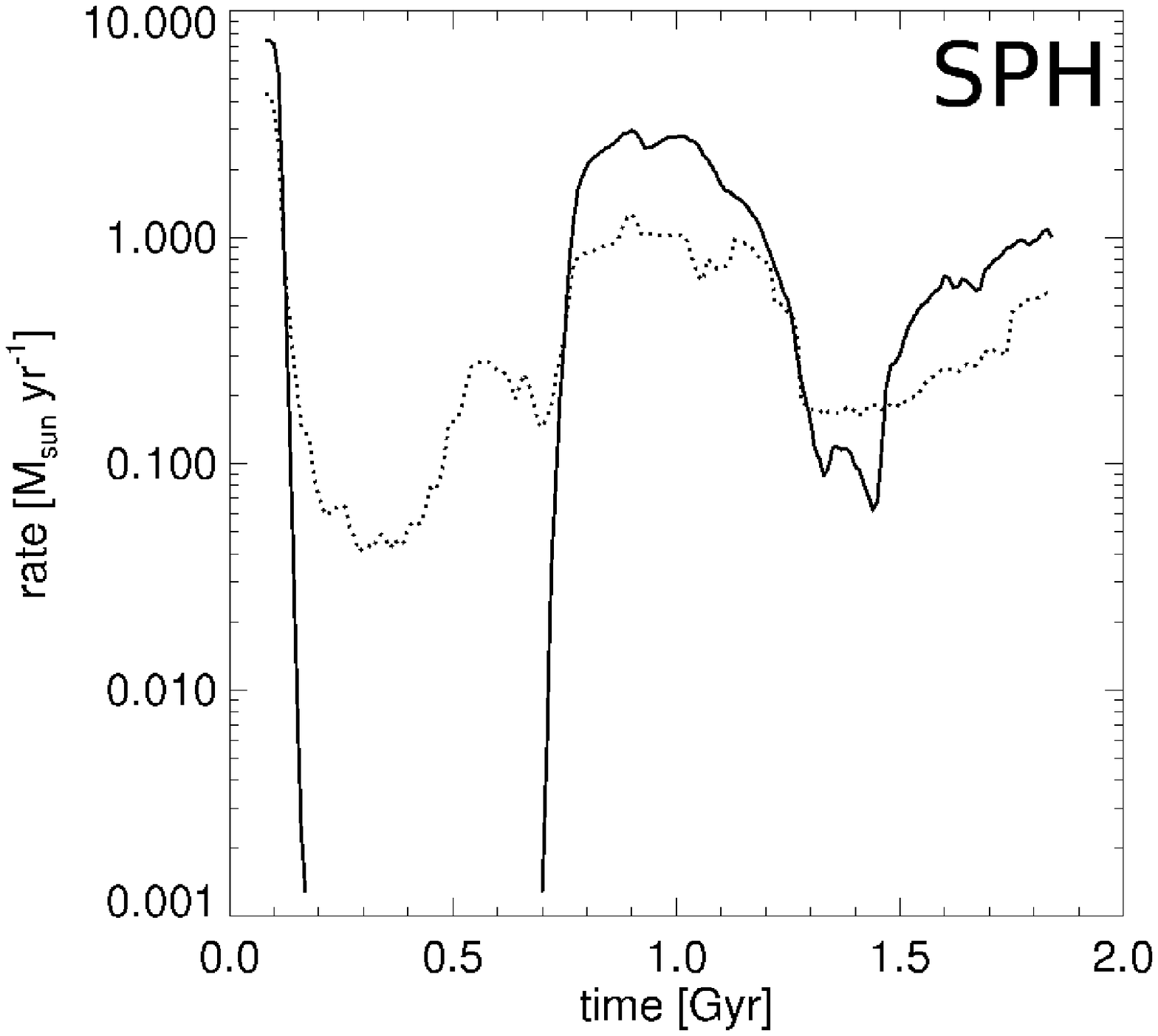,width=1.00\textwidth,angle=0}}
\end{minipage}
\begin{minipage}[b]{.49\textwidth}
\centerline{\psfig{file=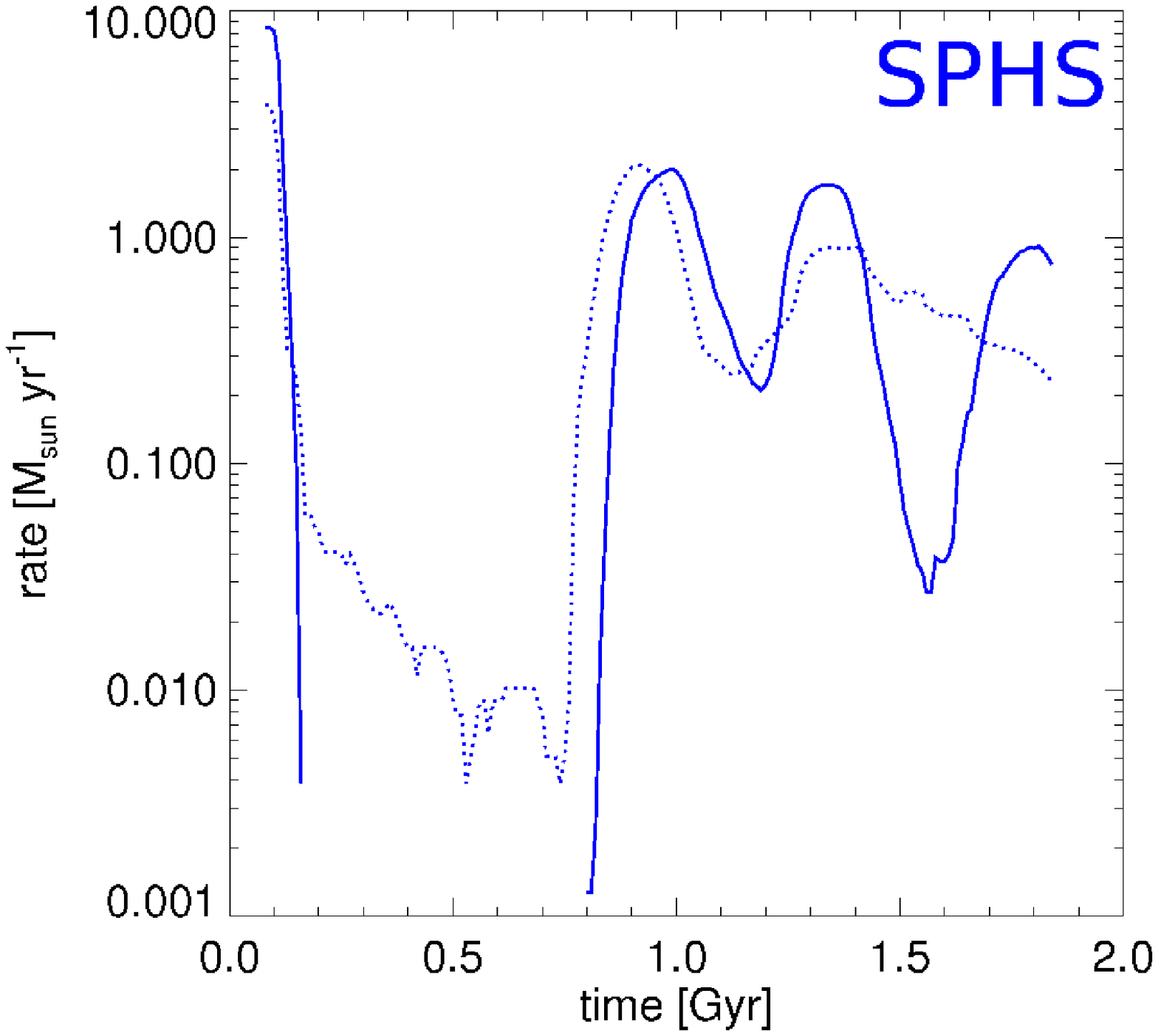,width=1.00\textwidth,angle=0}}
\end{minipage}
\caption{SFR and SMBH accretion histories for $N_{\rm gas} = 1 \times 10^6$, with SPH (left) and SPHS (right). Solid line is SFR, dotted is $\dot{M}_{\rm bh}$. The initial starburst has a slightly higher SFR in SPHS but the initial accretion rate is higher in SPH. Both rates then drop considerably for the next $\sim 1$\,Gyr with the SFR dropping to almost zero and the accretion rate by several orders of magnitude. During this period the SPH accretion rate remains higher than in SPHS by a factor of $\sim 10$. The second major starburst occurs later in SPHS than in SPH, has a lower SFR and drops off faster. However, in SPHS there are two subsequent major starbursts rather than just one in SPH. Again, the SMBH accretion rate approximately follows the shape of the SFR, with a small offset.}
\label{fig:sfrsmbh1e6}
\end{figure*}

The higher resolution runs constitute our fiducial comparison between the two methods. Gas surface density projections at $t = 1.2$ Gyrs are shown in Figure \ref{fig:snapshotinfall1e6}. Once again the initial starburst and feedback events are similar, although the initial feedback is slightly stronger in SPHS than in SPH, with the kicked gas reaching a maximum of $10^8$\,K vs. $2 \times 10^7$\,K and again expanding faster. At this resolution the peak of the star formation rate for this starburst is slightly higher in SPHS. Due to the stronger feedback, more of the gas is evacuated in the initial overpressurised bubble, and takes longer to make its way back inside the expanding shell. The accretion rate in SPHS is therefore reduced for the period immediately following the first starburst (between $\sim 0.2 - 0.7$\,Gyr) compared to SPH (see Figure \ref{fig:sfrsmbh1e6}).

Also at this higher resolution there are subsequent major starbursts in both the SPH and SPHS runs. The second large starburst occurs at $\sim 1$\,Gyr, and takes place slightly earlier in SPH than in SPHS. This time the second SPH starburst is more powerful than the SPHS one. We attribute this to the fact that immediately before the second starburst the SPH run saw multiple condensing gas clumps entering the central few kiloparsecs, which was not seen in SPHS. The expanding bubbles in the second starburst also differ noticeably between the two methods -- in SPH the boundaries of the cavities are thinner whereas in SPHS these outer `walls' are more mixed in with the surrounding gas (see Figure \ref{fig:snapshotinfall1e6}).

The star formation and accretion histories for the higher resolution runs can be seen in Figure \ref{fig:sfrsmbh1e6}. Most noticeable is the presence of three peaks in the SFR to later times in the SPHS run, compared to two peaks in SPH. Once again the accretion rate largely follows the SFR, although not precisely, and in particular we notice that whenever there is a strong dip in the SFR, the corresponding dip in the $\dot{M}_{\rm bh}$ is not as extreme. This is also true for the peaks, with the exception of the first peak of the later starburst/accretion events in SPHS at $t \sim 1$\,Gyr. We have already mentioned how the second SPH starburst is more powerful than its SPHS equivalent, and this is seen also in the SFR, with the SPH peaking at a higher rate and falling off more slowly.

From this point on our comparisons are made with the higher resolution set of runs. We outline a few specific differences that we wish to highlight between the results obtained with the SPH and SPHS methods.

\subsubsection{Clumps vs. filaments}\label{sec:clumps}

Of particular interest is the behaviour of the two methods just after $t = 1$\,Gyr. As we have already mentioned, the SPH run has started to see condensing clumps form out of the halo gas, which fall into the centre and form stars, giving rise to a powerful second starburst. As the bubble expands through the surrounding gas, there appears to be a sudden condensation of multiple dense clumps from the previously low-density gas. These clumps scatter off each other as they fall to the centre, but remain seemingly protected from the ambient gas through which they travel. The clumps condense to the point that they reach the polytrope (refer to Sections \ref{sec:cooling} \& \ref{sec:sf}) and therefore undergo star formation and subsequent feedback. The result is that the flow in the inner $\sim 100$\,kpc becomes very disordered, with continued asymmetric feedback events of varying strengths. The clumps that reach the central few kiloparsecs either add to or remain in orbit of the disc, although some are accreted.

In the SPHS run, however, this condensation of cold, dense clumps does not occur. Rather, the structures that form are filamentary in nature, forming at the intersection between bubbles, with gas infalling through the filaments to grow the disc (or to be accreted by the SMBH). The very few spherical overdensities that do begin to form are quickly disrupted by travelling through the ambient gas before they can reach the densities required to lie on the polytrope. This quite striking comparison can be seen clearly in Figure \ref{fig:snapshotinfall1e6}, where the structure formation and mode of feeding is clearly different between the two methods. In Figures \ref{fig:phasecutsSPH1.2} \& \ref{fig:phasecutsSPHS1.2} (second from top, left-hand panels), we isolate the overdense regions through a series of cuts in the phase diagram. The differences between the cold clumps (SPH) and filaments (SPHS) can clearly be seen. In Section \ref{sec:numericalclump} we discuss the reasons for these differences and demonstrate that the clumps are a numerical artifact.

\subsubsection{Disc formation}

\begin{figure*}
\begin{minipage}[b]{.49\textwidth}
\centerline{\psfig{file=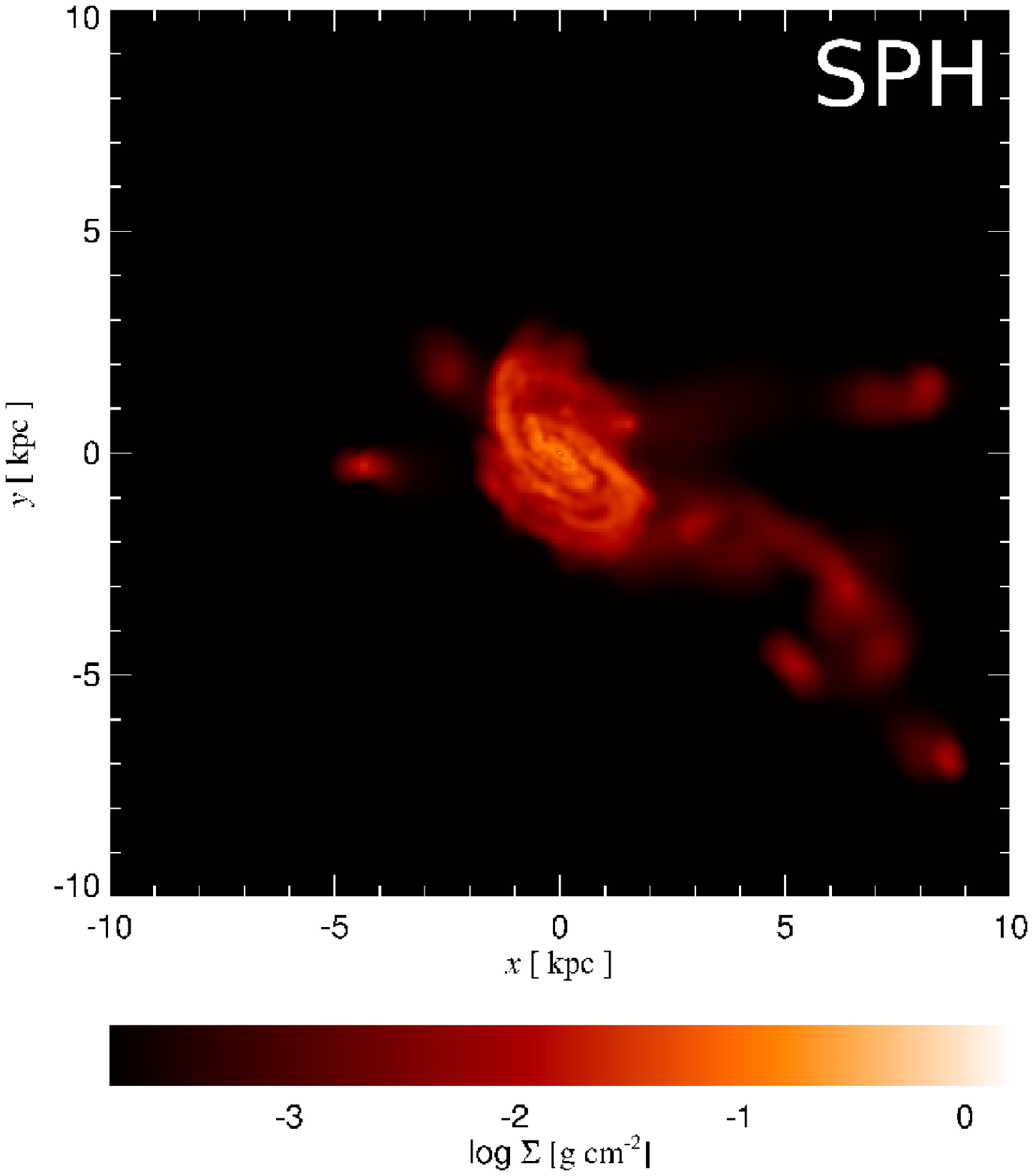,width=1.00\textwidth,angle=0}}
\end{minipage}
\begin{minipage}[b]{.49\textwidth}
\centerline{\psfig{file=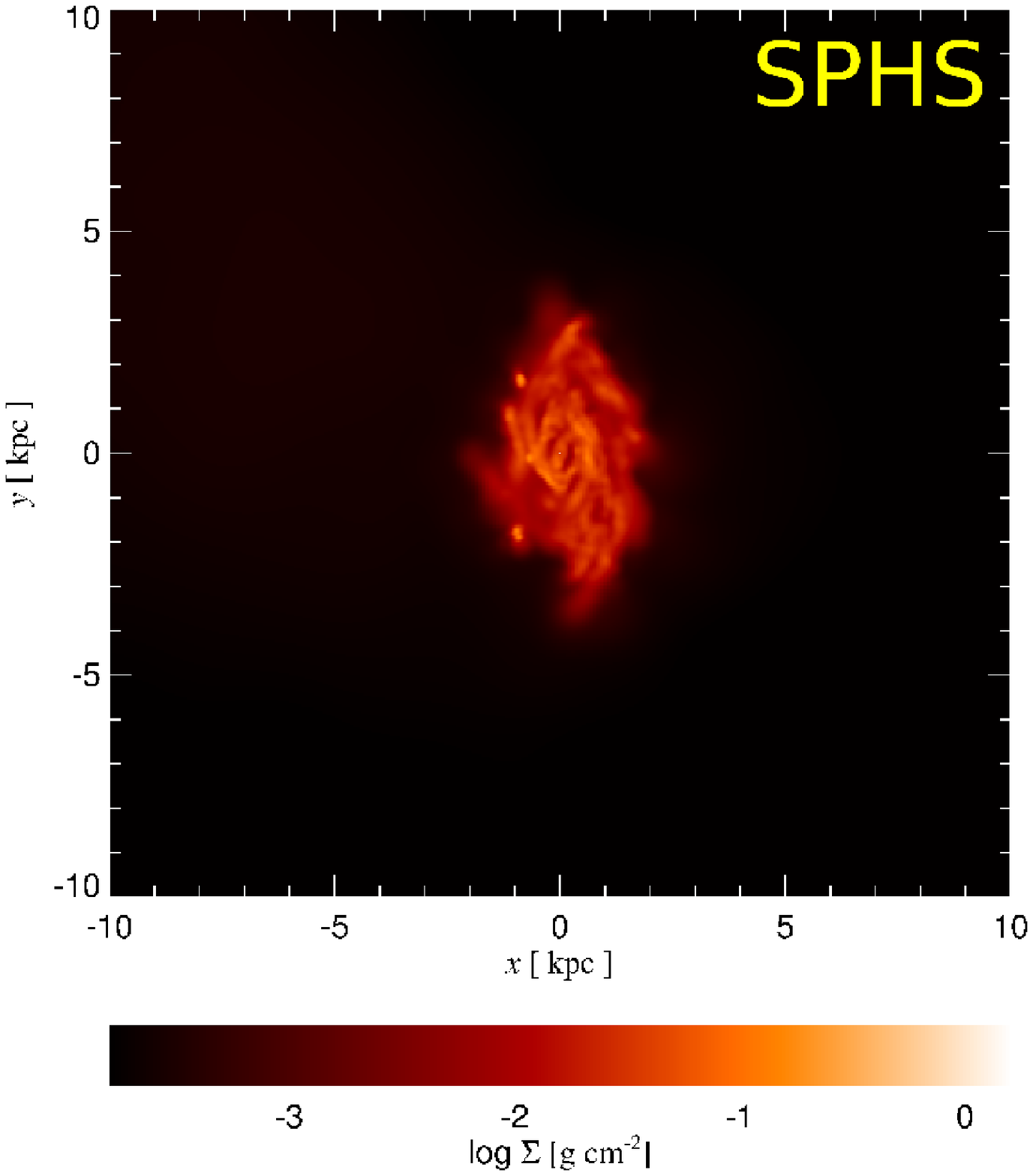,width=1.00\textwidth,angle=0}}
\end{minipage}
\caption{Projected surface density plots of the central disc that forms in the SPH-96-res1 (left) and SPHS-96-res1 (right) runs, at $t = 2$\,Gyrs. Both discs are of a similar size, although the stochastic mode by which they have grown is very dependant on the random orientations of the supernovae feedback, and they are therefore at different orientations despite the same initial velocity field for the gaseous halo. The SPH disc is surrounded by multiple clumps that formed in the halo whereas the clumps present in the SPHS case have formed from the disc itself.}
\label{fig:snapshotinfall1e6disc}
\end{figure*}

Figure \ref{fig:snapshotinfall1e6disc} shows the surface density projection of the gaseous disc that forms in the higher resolution SPH and SPHS runs. Both discs have a similar mass of $\sim 2 \times 10^9 \msun$, and are of a similar size, extending out to $\approx 2.5$\,kpc. These similarities are understandable given the identical initial condition used for both runs. However, the formation of the disc is strongly influenced by the asymmetries present in the star formation and feedback events, which eject hot bubbles in a variety of directions into the gaseous halo. It is therefore not surprising that the discs in the SPH and SPHS runs are at different orientations, as the star formation and feedback histories of the two are somewhat different. The stochastic nature of the feeding to the central regions is so significant that neither disc is in the preferred $x-y$ plane set by the angular momentum of the initial conditions. The two discs do in fact start out in the $x-y$ plane as they are initially formed in the first starburst, but undergo precession and midplane rotation as they are torqued by subsequent infalling gas at a variety of different orientations.

One important distinction between the discs in the two methods is the presence of the clumps orbiting outside the disc in the SPH case, which are not there in SPHS. This relates to the clumpy vs. filamentary mode of feeding the disc as discussed above in Section \ref{sec:clumps}. There are clumps \emph{within} the disc in the SPHS run, but they have formed out of the disc itself through gravitational instability, rather than being formed further out in the halo and migrating to the centre.

\subsubsection{Galaxy morphology and stellar content}

\begin{figure*}
\begin{minipage}[b]{.49\textwidth}
\centerline{\psfig{file=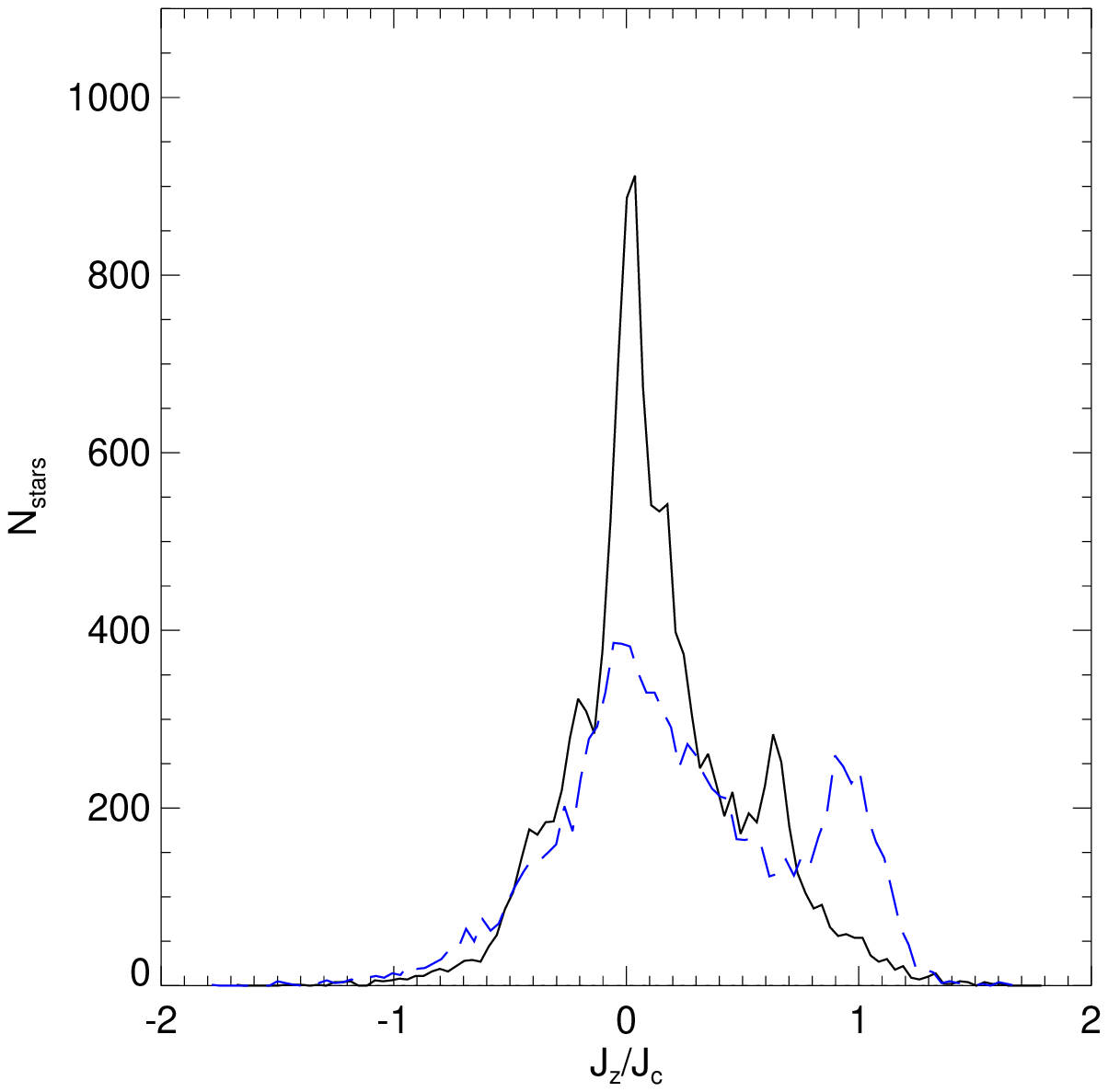,width=1.0\textwidth,angle=0}}
\end{minipage}
\begin{minipage}[b]{.49\textwidth}
\centerline{\psfig{file=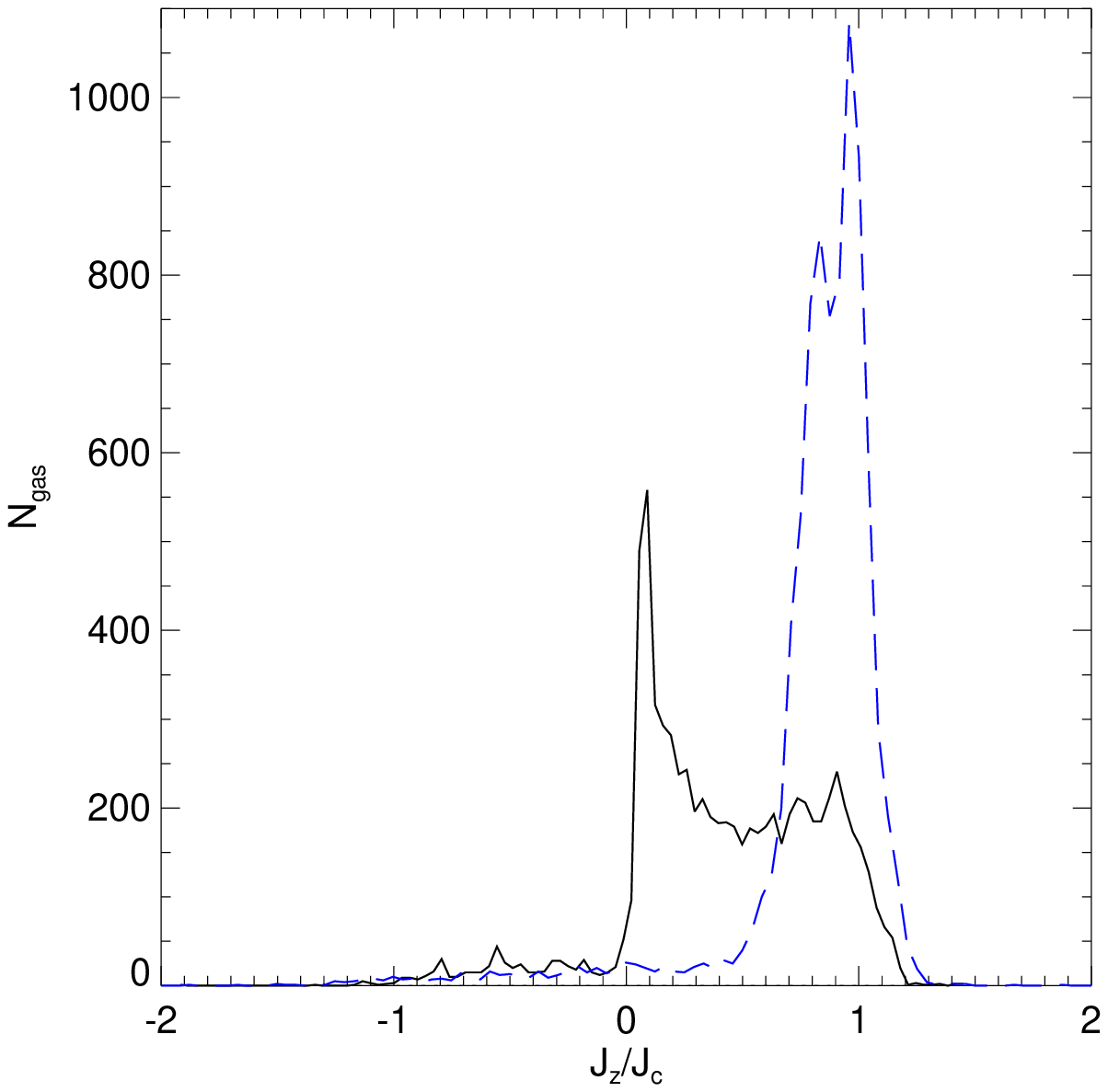,width=1.0\textwidth,angle=0}}
\end{minipage}
\begin{minipage}[b]{.49\textwidth}
\centerline{\psfig{file=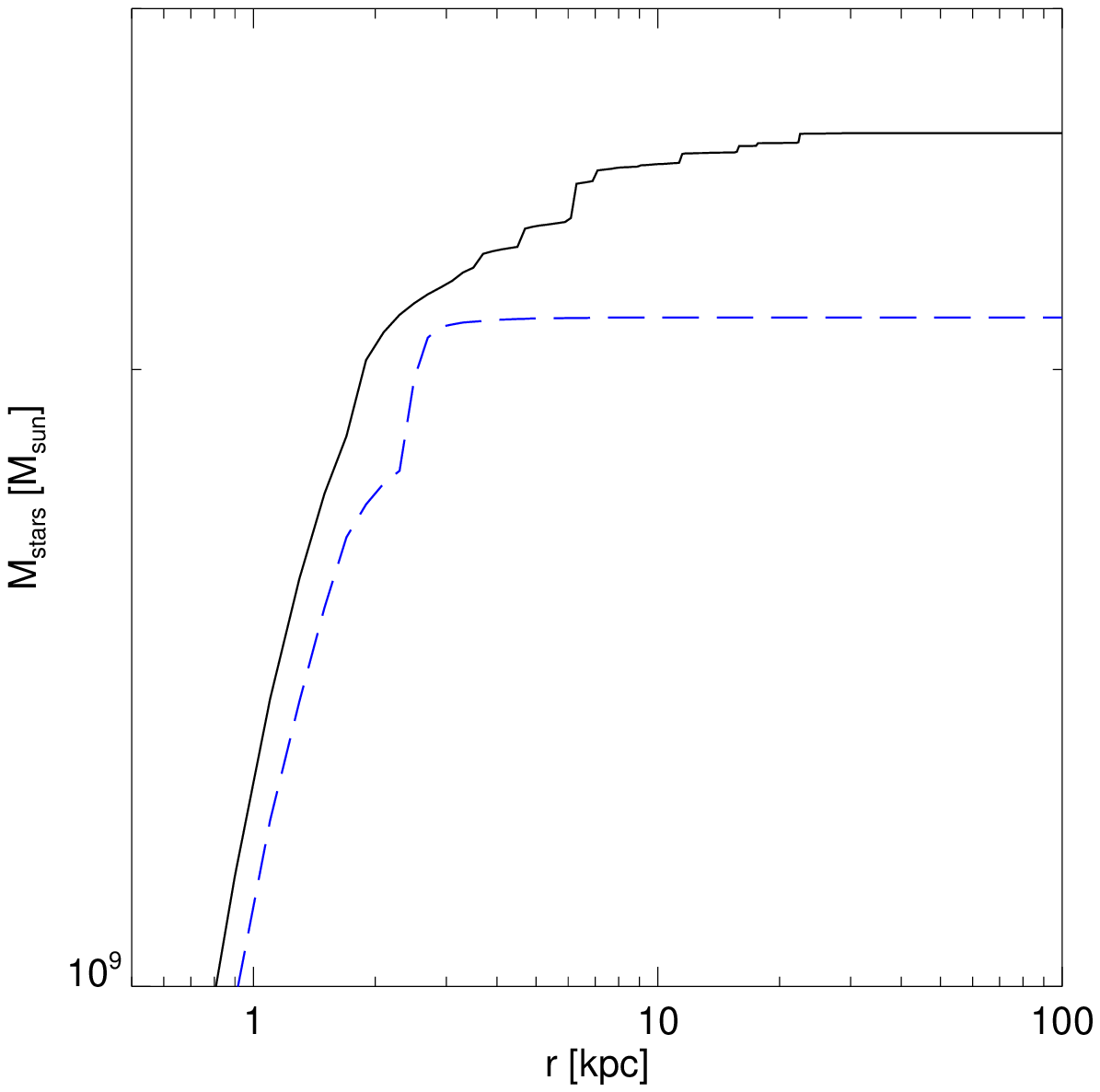,width=1.0\textwidth,angle=0}}
\end{minipage}
\begin{minipage}[b]{.49\textwidth}
\centerline{\psfig{file=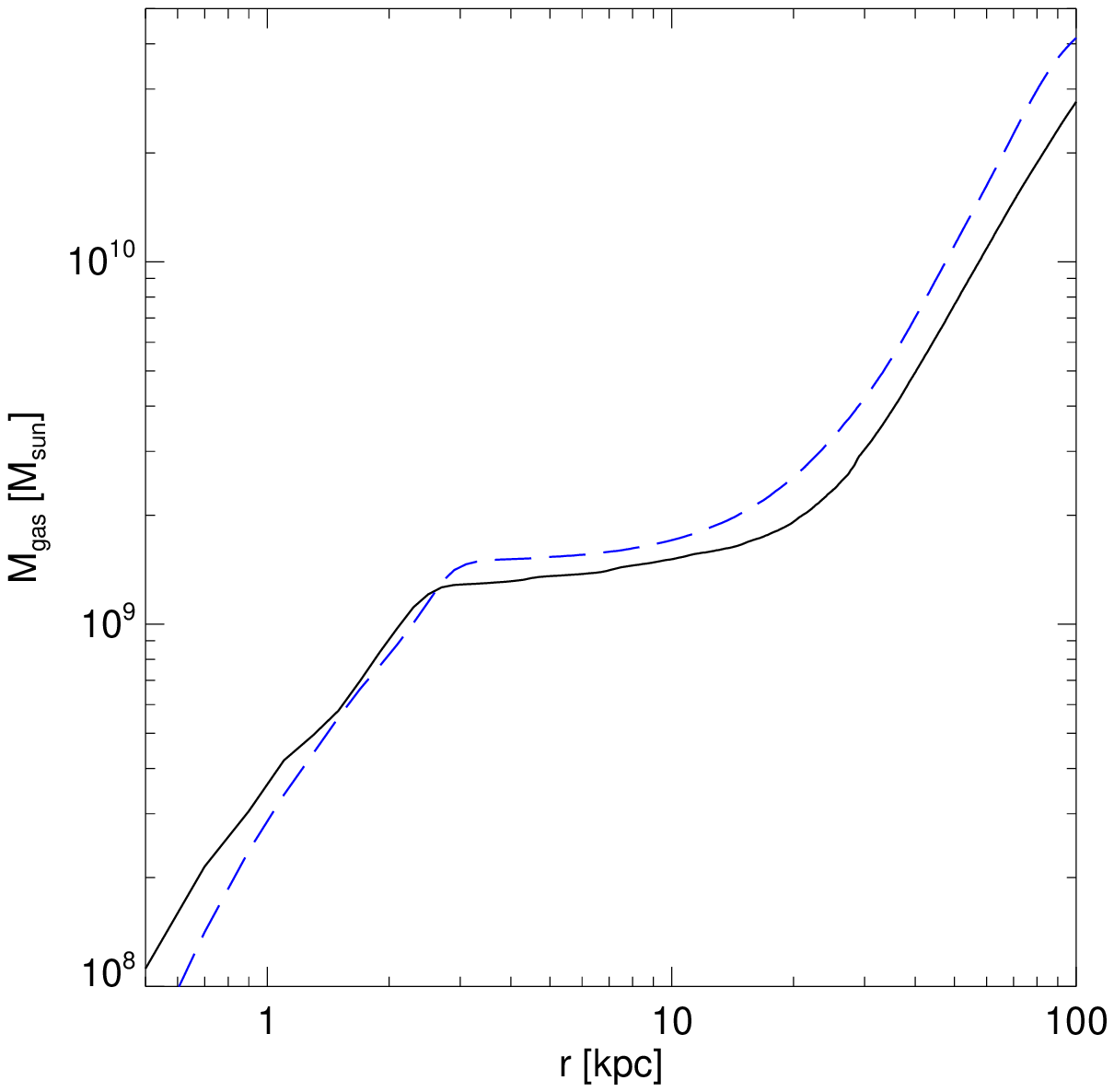,width=1.0\textwidth,angle=0}}
\end{minipage}
\caption[]{Plots of the kinematic `disc/bulge' ratio at $t = 2$ Gyr for the inner 8 kpc in the stars (top left) and the gas (top right) together with the enclosed mass in stars (bottom left) and gas (bottom right) out to 100 kpc for the SPH-96-res1 (black solid) and SPHS-96-res1 (blue dashed) simulations. The SPHS run has a more disc-like morphology inside the central 8 kpc, with the $J_z/J_c \simeq 1$ signal (a `disc') being a factor of $\approx 4$ higher than SPH in the stars and a factor of $\approx 5$ higher than SPH in the gas. The SPH run shows a stronger peak at lower angular momentum with a $J_z/J_c \simeq 0$ signal (a `bulge') that also corresponds to more randomised orbits than in SPHS.}  
\label{fig:diskbulge}
\end{figure*}

When we look in more detail at the kinematics of the gas and stars in the inner region, we find that the morphologies of both the stars and the gas actually differ by considerably more than a visual inspection of the disc (see Figure \ref{fig:snapshotinfall1e6disc}) would suggest. In particular, the SPH-96-res1 run shows a larger amount of low angular momentum material within the central $8$ kpc than is present in the SPHS equivalent. This can be seen clearly in Figure \ref{fig:diskbulge}, which shows plots of the $J_z/J_c$ ratio (often referred to as the `disc/bulge' ratio), where $J_z$ is the magnitude of the angular momentum vector along the rotation axis of the inner $2$ kpc of each disc, and $J_c$ is the magnitude of the angular momentum for a circular orbit at a radius $r$, namely $J_c = r v_c$, where $v_c^2 = GM(r)/r$ is the circular velocity from the enclosed mass of all species (gas + stars + dark matter + SMBH). This ratio was calculated for each star and for each gas particle and the histograms for both shown in the Figure (top left and right respectively). It is clear from these plots that the SPHS run shows a kinematically more pronounced disc than SPH, particularly in the gas but also in the stars as well. The many cold clumps that are formed in SPH provide the inner $8$ kpc with a greater amount of low angular momentum material (gaseous and stellar) as well as a greater randomness of orbits as they fall in. The top right plot tells us further that the gas disc seen in SPH in Figure \ref{fig:snapshotinfall1e6disc} is more disturbed and warped compared to the SPHS one, being strongly torqued out of its plane and heated by the infalling gas clumps at different orientations. We note that since it is this gas that tends to lie on the polytrope and is therefore `star-forming', the stars that form here will likely inherit the $J_z/J_c$ ratio seen for the gas in the top right plot. This means that to later times (e.g., $3-4$ Gyr), the stellar kinematic disc/bulge ratio will likely become even more disparate between the SPH and SPHS simulations, with SPH favouring a bulge and SPHS favouring a disc.

Also in Figure \ref{fig:diskbulge} we see a plot of the enclosed mass in stars with radius (bottom left) and enclosed mass of gas with radius (bottom right), showing that the stellar content of the galaxies in SPH and SPHS differ considerably outside of $\sim 2$ kpc. The SPH run has a greater amount of mass in stars at all radii, and in particular further out in the halo, where the cold clumps are forming stars out to $20-30$ kpc. In the gas the profiles are similar, but the SPHS run has a greater mass of gas at large radii in the hot halo.

\subsection{Features in $\rho-T$ space}

In order to follow the relevant behaviour and understand evolution of the simulations better, we analyse a number of aspects of them in terms of their locations on a phase diagram $\rho$ vs. $T$. We isolate a few key regions and follow the behaviour of the gas that inhabits these regions on the phase diagram. The corresponding plots for this section are Figures \ref{fig:phasecutsSPH1.2} \& \ref{fig:phasecutsSPHS1.2}, in which the phase diagrams are plotted along with the projected surface density and projected temperature for the isolated phase regions.

\subsubsection{Isothermal contraction phase}\label{sec:isothermal}

As the gas cools out of hydrostatic equilibrium, it experiences a variable cooling rate that is set by the combined cooling curve of \cite{KWH1996} and \cite{MashchenkoEtal2008} used in our simulations. At $T = 10^4$\,K (the switchover between the two curves) the cooling becomes inefficient, and so gas tends to `pile up' at around this temperature. We have isolated a region on the phase diagram (2nd from top) that corresponds to gas that has cooled to this temperature and is condensing, moving nearly isothermally across the marked region to higher densities and eventually onto the polytrope. This we refer to as an `isothermal contraction phase', and has particular relevance to the many spurious clumps that are seen in the SPH runs, as it marks the region on the phase diagram that they inhabit. Although still present in the higher resolution SPHS runs, this region is significantly less occupied, and does not extend as far to low densities as in the SPH case.

\subsubsection{Polytrope phase}

As mentioned in Sections \ref{sec:cooling} \& \ref{sec:sf}, the diagonal line shown on the phase diagram corresponding to a polytropic EQS with an adiabatic index of $4/3$ functions as a dynamic pressure floor, ensuring that the Jeans mass is always resolved in the gas. The polytrope `fills in' from gas that either cools directly onto it or from gas that reaches it through the isothermal contraction phase discussed above. Once on the polytrope, the gas can move along it to higher densities (and higher temperatures as per the EQS) and of course can move off it either by an increase in temperature for a given density, or by a decrease in density for a given temperature. Gas can also leave the polytrope by being converted into stars, which is allowed to occur beyond the fixed threshold of $10^2$ atoms cm$^{-3}$, denoted by the green dotted line. The presence of gas on the polytrope past this fixed threshold is therefore transitory -- it fills in and subsequently disappears as stars form.


\begin{figure*}
\centerline{\psfig{file=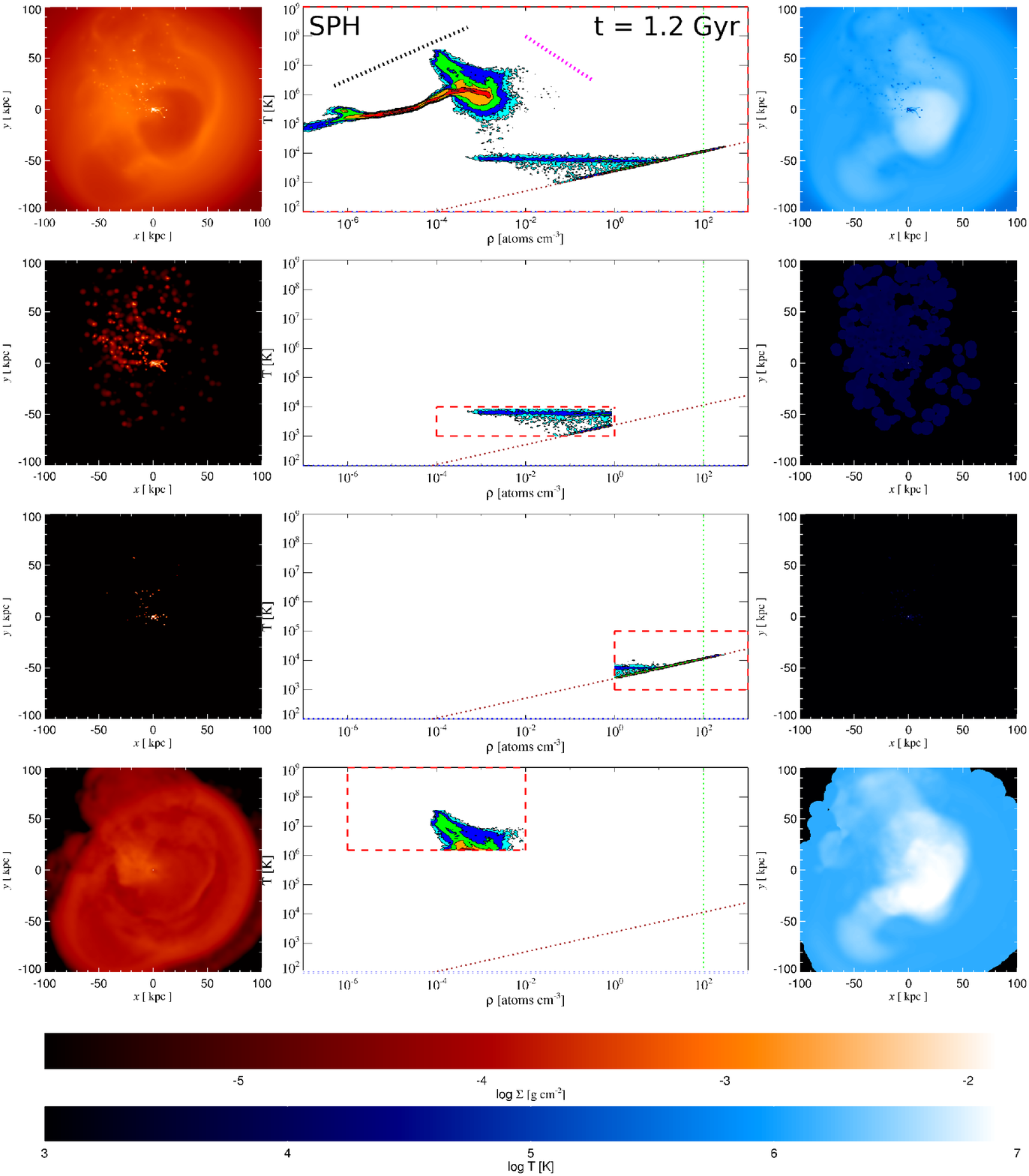,width=1.0\textwidth,angle=0}}
\caption[]{Projected surface density plots (left), phase diagrams (middle) and projected temperature plots (right) in SPH-96-res1 at $t = 1.2$\,Gyr. The white dot at $(0,0,0)$ marks the SMBH. The blue dotted line indicates the cooling floor, while the green and brown dotted lines mark the fixed SF threshold and polytropic pressure floor respectively. The two thicker diagonal lines indicate the standard Sedov-Taylor solution (magenta) and the post Sedov-Taylor adiabatic expansion (black) that fit the orientation of the feedback peaks and their evolution in time respectively. The red outlined boxes indicate the regions of the phase diagram being plotted. Contours are binned logarithmically, starting at the minimum value for $\rho$ or $T$ and increasing with a factor of $d \rho / \rho = dT/T = 0.1$. The colours correspond to the number of gas particles, with cyan representing individual particles and subsequent colours going up in factors of 5. In this simulation we see a great deal of structure, in the form of multiple overdense clumps, distributed over a large part of the computational domain and lying within the `isothermal contraction' region of the phase diagram. We also see a broadened spike along the Sedov-Taylor gradient that corresponds to many different supernova feedback events.}
\label{fig:phasecutsSPH1.2}
\end{figure*}

\begin{figure*}
\centerline{\psfig{file=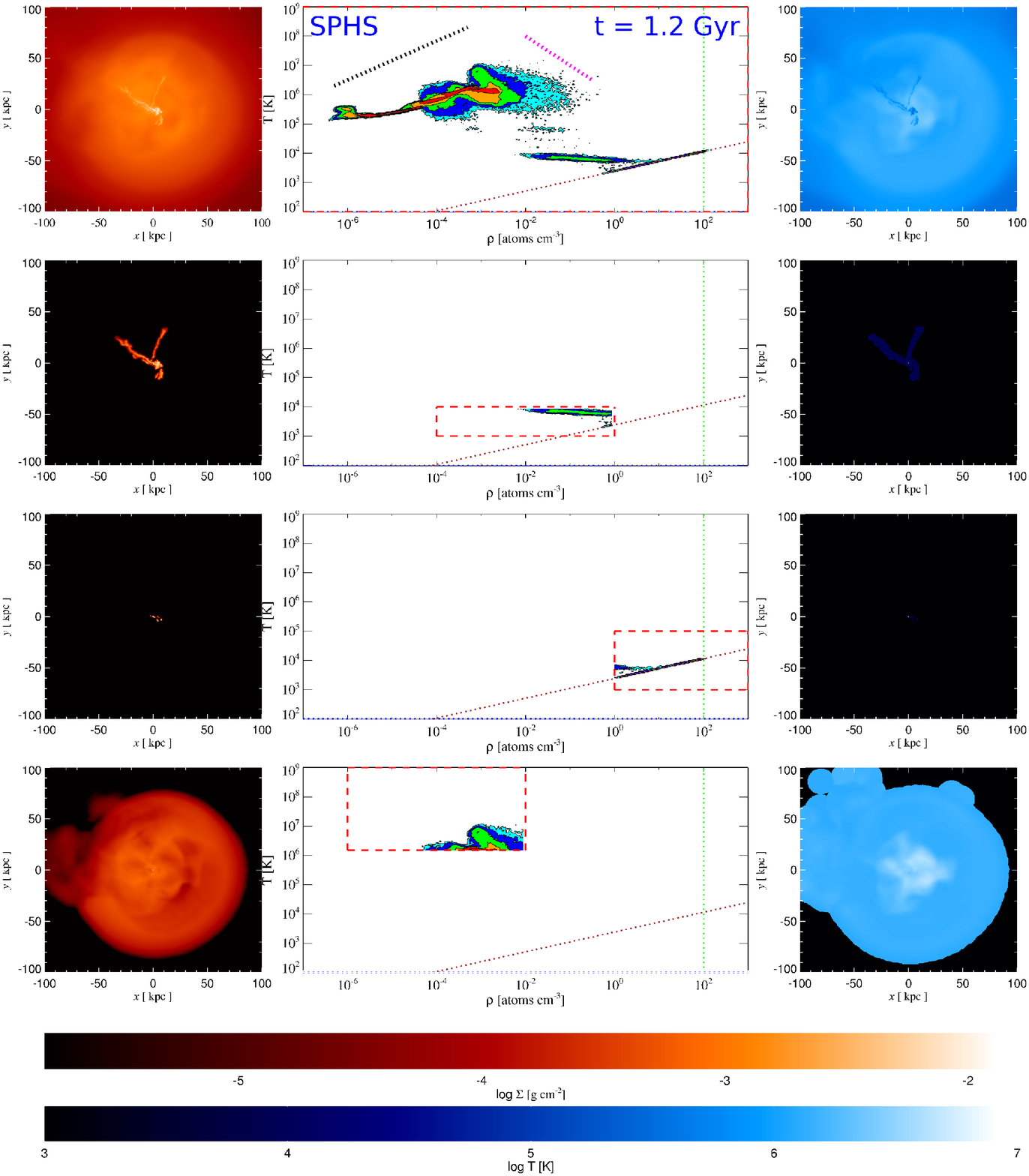,width=1.0\textwidth,angle=0}}
\caption[]{Projected surface density plots (left), phase diagrams (middle) and projected temperature plots (right) in SPHS-96-res1 at $t = 1.2$\,Gyr. Linestyles etc. are as per Figure \ref{fig:phasecutsSPH1.2}. There are striking differences in the structures that have formed compared to the SPH case (Figure \ref{fig:phasecutsSPH1.2}) -- in the latter we saw multiple clumps whereas here we see one or two filaments. Again, the structure occupies the isothermal contraction region of the phase diagram, with the gas flowing down the filaments to feed the disc and form stars. Similar to the SPH case, we see a broadened spike along the Sedov-Taylor gradient corresponding to multiple feedback events, but reduced in magnitude.}
\label{fig:phasecutsSPHS1.2}
\end{figure*}

\subsubsection{Hot bubble phase}\label{sec:bubbles}

The filling in of the polytrope beyond the fixed density threshold for SF is often followed by an ejection event that pushes gas into the `overpressurised bubble' section of the phase diagram. Within this region there we can identify both the $\rho-T$ trend for a given bubble, as well as the evolution of this trend with time, through analytic arguments. To start with, the similarity solution for a Sedov-Taylor blast wave due to an energy deposition $E$ in a uniform medium \citep{Sedov1959} is given by:
\begin{equation}
r(t) \propto \left(\frac{E t^2}{\rho_{\rm ISM}}\right)
\end{equation}
where $\rho_{\rm ISM}$ is the average (constant) density of the surrounding medium. We can define an overdensity parameter $\delta_s \equiv \rho_s/\rho_{\rm ISM}$ for the shocked gas, which under the assumption of a strong adiabatic shock, is a constant -- given by $\delta_s \simeq 4$. The shock velocity, $v_s = \text{d}R/\text{d}t$ is therefore:
\begin{equation}
v_s \propto \frac{2}{5} E^{1/5} (4 \rho_s)^{1/5} t^{-3/5} = \frac{2r}{5t}
\end{equation}
which we can write as:
\begin{equation}
v_s \propto E^{1/2} r^{-3/2} \rho_s^{-1/2}
\end{equation}
The post-shock temperature for an adiabatic shock with velocity $v_s$ is:
\begin{equation}
T_s \propto \frac{\mu m_p v_s^2}{k_B}
\end{equation}
which gives us:
\begin{equation}
T_s \propto E r^{-3} \rho_s^{-1}
\end{equation}
The trend $T \propto \rho^{-1}$ is shown on the top phase plot (dotted magenta line) in Figures \ref{fig:phasecutsSPH1.2} \& \ref{fig:phasecutsSPHS1.2}. The shape of this feature remains constant but evolves along the phase diagram as the bubbles expand. Such evolution can be described analytically by a post Sedov-Taylor solution for adiabatically-expanding hot gas. If we assume that the expansion is sufficiently fast that radiative cooling can be neglected, the equation of state is $P \propto \rho^\gamma$, with $\gamma = 5/3$. We can write this in terms of the sound speed by noting that:
\begin{equation}
c_s^2 = \frac{\gamma P}{\rho}
\end{equation}
and therefore that:
\begin{equation}
P = \frac{\rho k T}{\mu m_p}
\end{equation}
which we equate to obtain:
\begin{equation}
T \propto \rho^{2/3}
\end{equation}
for the time evolution of the bubble. This trend is also plotted in the top phase diagram (dotted black line) in Figure \ref{fig:phasecutsSPH1.2} \& \ref{fig:phasecutsSPHS1.2}.

Even though the analysis described above corresponds to the solution for a uniform medium, it fits the supernovae ejecta in our simulations remarkably well\footnote{to see the time evolution clearly the reader is directed to \url{http://www.phys.ethz.ch/~ahobbs/movies.html}}. The reason for this is that the two departures from the standard Sedov-Taylor solution that are present in our model exert competing effects on the $\rho-T$ profile of our hot bubbles. Firstly, we have a non-uniform medium, where $\rho$ falls off according to equation \ref{eq:rho}. This has the effect of reducing the velocity fall-off as the shell propagates and sweeps up mass. Secondly, we have an inwardly-directed velocity field as the gas cools and inflows under the influence of the gravitational potential, which has the effect of enhancing the velocity fall-off. The infall velocity arises self-consistently from the potential, which is a function of the density profile $\rho(r)$. The two departures from standard Sedov-Taylor therefore balance out, and the majority of the supernovae-driven bubbles behave according to the analytical arguments above (with small additional departures due to radiative cooling and the modification of the density profile due to previous supernovae explosions).

\subsection{Numerical clump formation}\label{sec:numericalclump}

\begin{figure*}
\centerline{\psfig{file=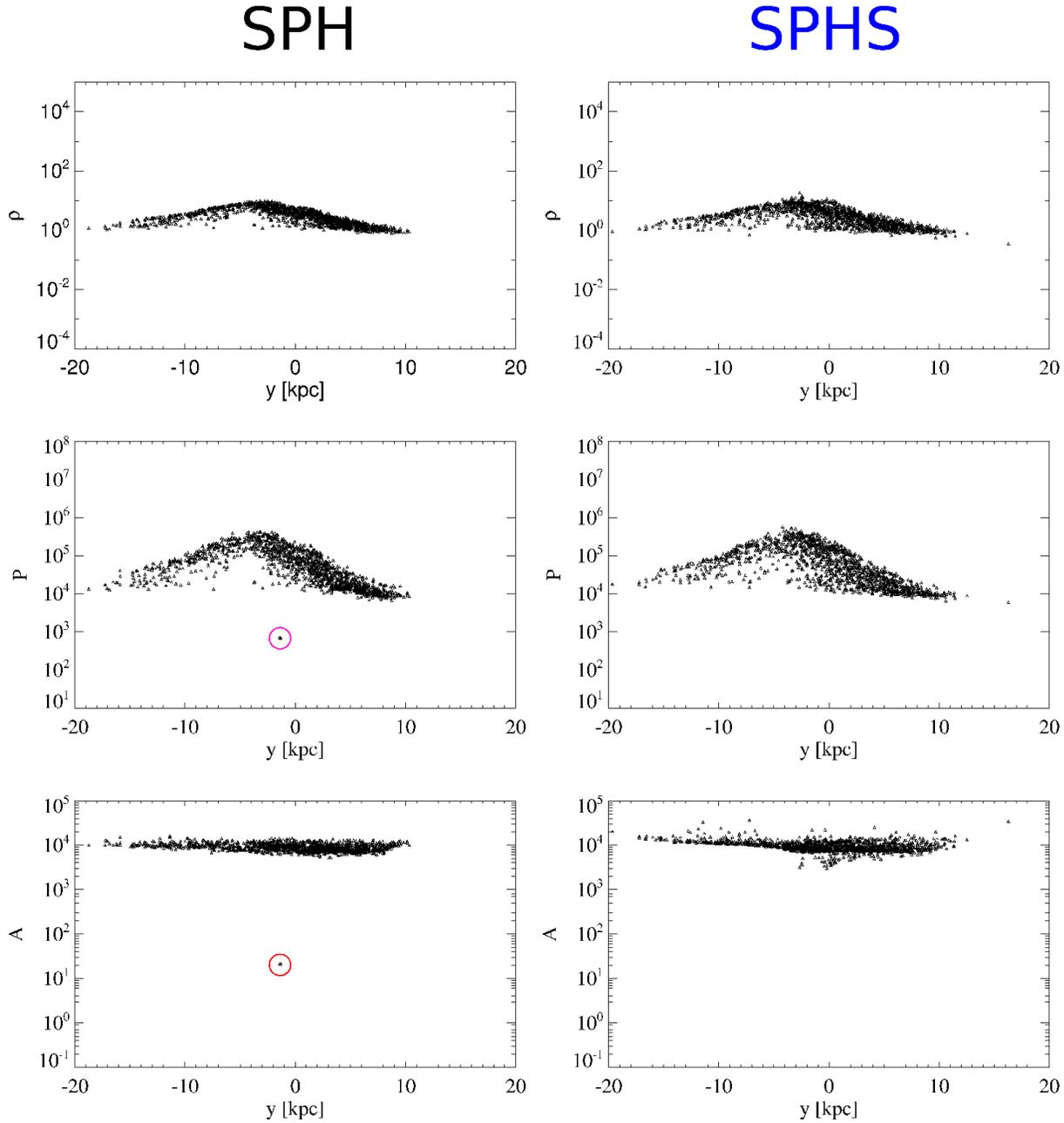,width=1.0\textwidth,angle=0}}
\caption[]{Plots of density (top), pressure (middle) and entropy (bottom) for the most gas-rich overdense clump in the SPH-96-res1 run, at $t = 0.63$ Gyr. The gas that formed the clump was identified and tracked back earlier in the simulation to before it formed. On the left-hand side are the plots showing the evolution of this particular clump with the SPH method, while on the right-hand side the plots show the evolution of the clump with the SPHS method (although starting from the same SPH-96-res1 snapshot from which the clump was taken -- at $t = 0.4$\,Gyr). Each property ($\rho$, $P$, $A$) is plotted in the centre of mass frame of the clump gas at a time in its evolution where a density peak that occurred just previously has been smoothed out. This density peak had a corresponding entropy dip at the same location. When evolved with SPHS, the smoothing of the density peak coincides with a diffusion of the entropy, removing the dip and allowing the pressures to remain smooth; however, when exactly the same initial condition is evolved with SPH the entropy dip remains (red circle), driving an equivalent dip in the pressures (magenta circle). This central pressure dip drives the contraction of the gas and the subsequent formation of the clump. For the full evolution and a clear picture of how this occurs the reader is directed to the corresponding movie at \url{http://www.phys.ethz.ch/~ahobbs/movies.html}.}
\label{fig:clumptrackingSPH}
\end{figure*}

\begin{figure*}
\centerline{\psfig{file=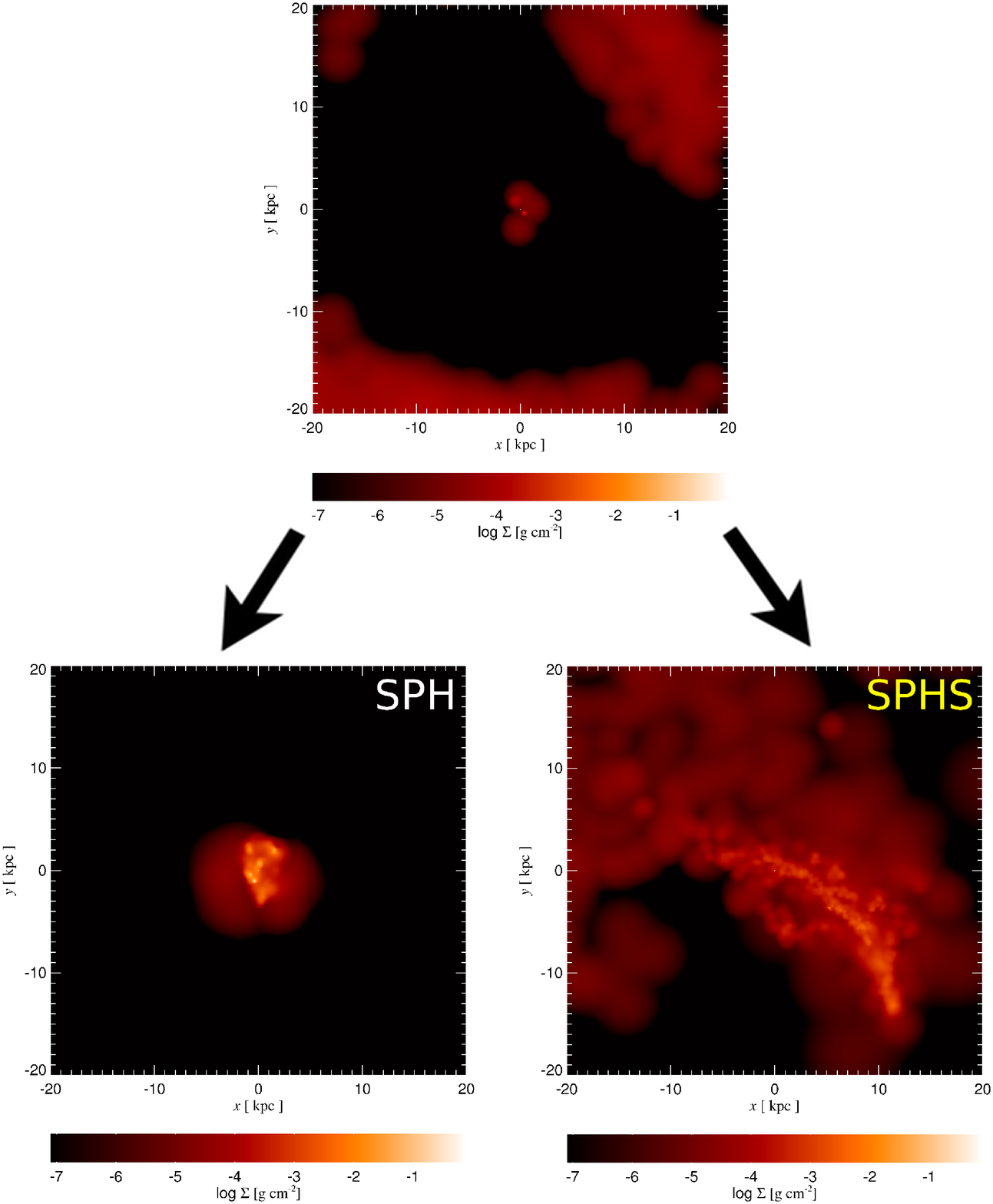,width=1.0\textwidth,angle=0}}
\caption[]{Visualisation of the numerical experiment discussed in Section \ref{sec:numericalclump} and Figure \ref{fig:clumptrackingSPH}. The initial condition (top plot) at $t = 0.4$\,Gyr is identical (taken from the SPH simulation of an earlier state of the gas in one of the bound clumps) but was evolved both with SPH (bottom left) and with SPHS (bottom right) to a time of $t = 1$\,Gyr. The differences are striking; the SPH evolution of the gas has formed a near spherical, dense clump while the SPHS evolution shows a stretched out filament of lower density. The point at which the two diverge is shown in Figure \ref{fig:clumptrackingSPH}, where a pressure discontinuity at the centre of the merging gas in the SPH case causes an artificial contraction that leads to the formation of the clump; this feature is not present in the SPHS case.}  
\label{fig:clumpfinalSPHS-SPH}
\end{figure*}

So far, it is clear that the presence of the overdense clumps is a feature of the SPH method, but the question remains as to what causes them. The fact that an identical simulation run with SPHS (CS kernel with 96 neighbours) does not yield these structures suggests that the prevention of multivalued fluid quantities plays a role in avoiding their formation, as this is the main difference between the methods when the kernel and neighbour number are not sufficient to have improved force accuracy \citep[refer to][]{2012MNRAS.tmp.2941R}. In this section, we show that this is indeed the case; it is the removal of pressure blips in an otherwise smooth flow that prevents the condensation of the clumps.

To identify bound gaseous clumps, we ran the Amiga halo finder \citep{AHF1, AHF2} on each simulation output. We focus here on the most gas-rich clump in the SPH run. Tagging the clump particles, we traced the evolution of the clump back in time to its initial formation. This particular clump forms from the merging of three distinct regions of gas, one of which has a significantly higher density and lower temperature than the other two. As the three regions merge, a small density peak forms with a corresponding entropy dip caused by gas cooling in the peak. The flow at this point is strongly shearing and the density peak rapidly shears away. By contrast, however, the entropy dip remains due to the lack of multiphase mixing in SPH. The presence of an entropy dip with no corresponding density peak drives a pressure dip at the centre of the merging gas. This can be seen on the left-hand side of Figure \ref{fig:clumptrackingSPH}, where the entropy and pressure dips are marked by the red and magenta circles, respectively. The central pressure dip drives an inward collapse and the formation of a bound clump (see Figure \ref{fig:clumpfinalSPHS-SPH}). 

In order to be sure that this multivalued pressure problem is not present in the new SPHS method, we took the starting conditions from the SPH run at $t = 0.4$\,Gyr -- just before the `clump gas' began to merge -- and ran it with SPHS (CS kernel, 96 neighbours). Once again, we tracked the evolution of the gas which, in the SPH case, ended up in the clump. The initial merging occurred in exactly the same way, but as the density peak sheared away, so too did the dip in the entropy. As a result, there was no pressure dip and therefore no initial seed for collapse. This can be seen on the right-hand side of Figure \ref{fig:clumptrackingSPH}, where the features marked by the circles on the left-hand side are not present.

Figure \ref{fig:clumpfinalSPHS-SPH} shows the state of the same `clump-identified' gas in the SPH and the SPHS cases. The former, as a result of the `pressure-dip seed', has collapsed to form a near-spherical clump with multiple overdensities. In the SPHS run, however, the gas has been drawn out into a filament, or stream, having not been artificially pulled inwards due to a central dip in the gas pressure.

The filament/stream seen in Figure \ref{fig:clumpfinalSPHS-SPH} for the SPHS evolution of the merging clump shows individual density peaks forming along the central axis -- however, at this resolution these are dissimilar structures from the clumps seen in SPH, as they are transient, being mixed in with the surrounding gas before they can form stars. In the next Section we discuss the formation of bound clumps within the filament at higher resolution, which form due to the density enhancement created as two or more evacuated bubbles intersect.

\section{`Resolved' clump formation with SPHS: fragmentation of overdense filaments}\label{sec:fullpower}

We now move on from the comparison between the two methods at identical spatial resolution to a simulation that employs the full SPHS algorithm: in addition to the dissipation for advected fluid quantities, we also have improved force accuracy through the use of the HOCT4 kernel with 442 neighbours. We perform this simulation at significantly higher resolution than our fiducial comparison runs, reaching a gas particle mass of a few $\times 10^4 \msun$ (see Table 1).

The evolution of the higher resolution simulation is globally similar to those at lower resolution -- an initial central starburst sends a strong Sedov-Taylor bubble into the surrounding gas, which shuts off star formation and reduces SMBH accretion until such time as the gas makes its way back to the central regions. The subsequent starbursts are then highly asymmetric, with multiple cavities created from individual supernovae feedback events. Again, there is no sign of the many artificial clumps condensing from the ambient halo gas, and the structure that forms as the gas cools is filamentary in nature, caused by the merging of the outer walls of two or more of the feedback-driven bubbles.

\begin{figure}
\begin{minipage}[b]{.49\textwidth}
\centerline{\psfig{file=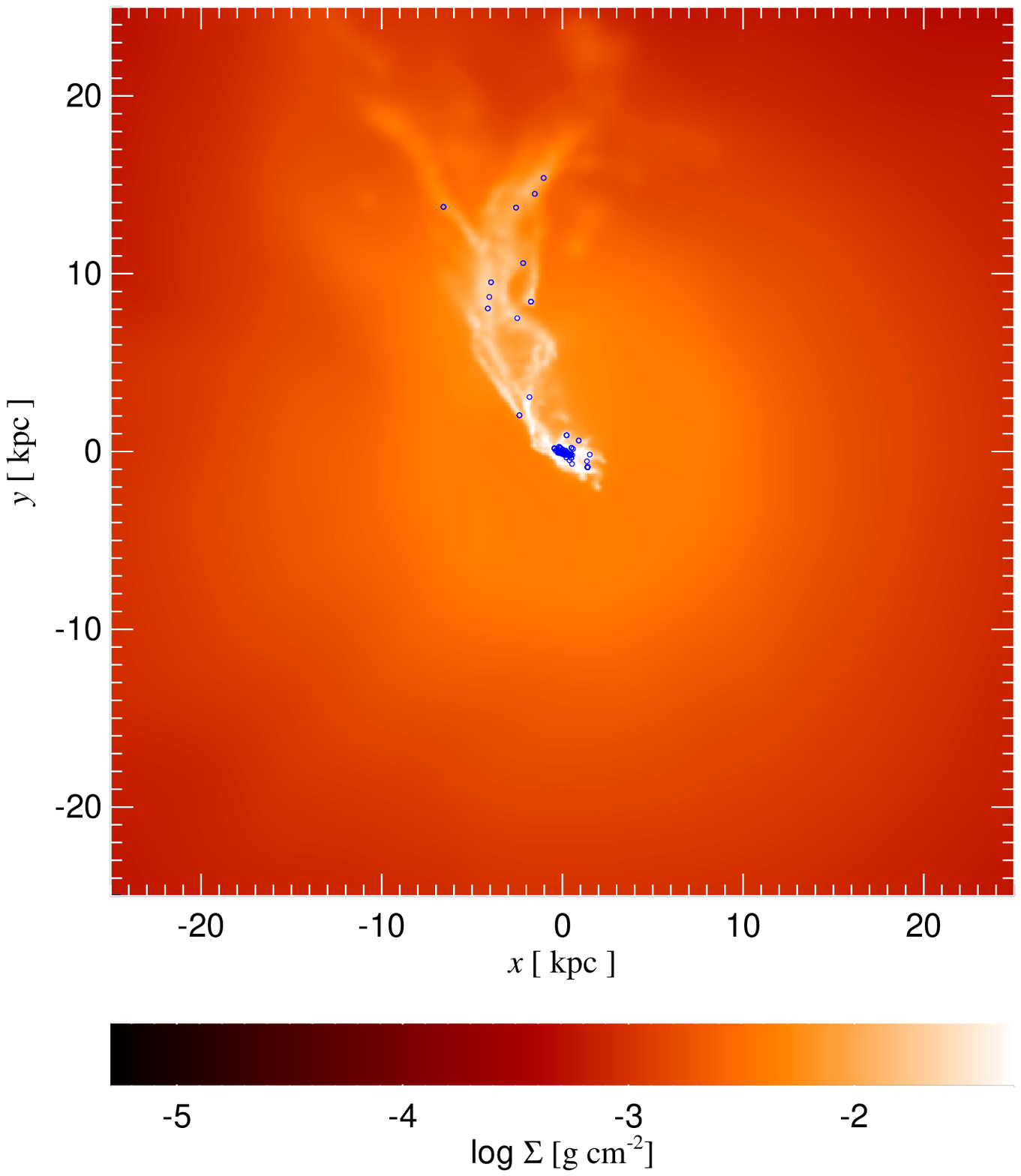,width=1.0\textwidth,angle=0}}
\end{minipage}
\begin{minipage}[b]{.49\textwidth}
\centerline{\psfig{file=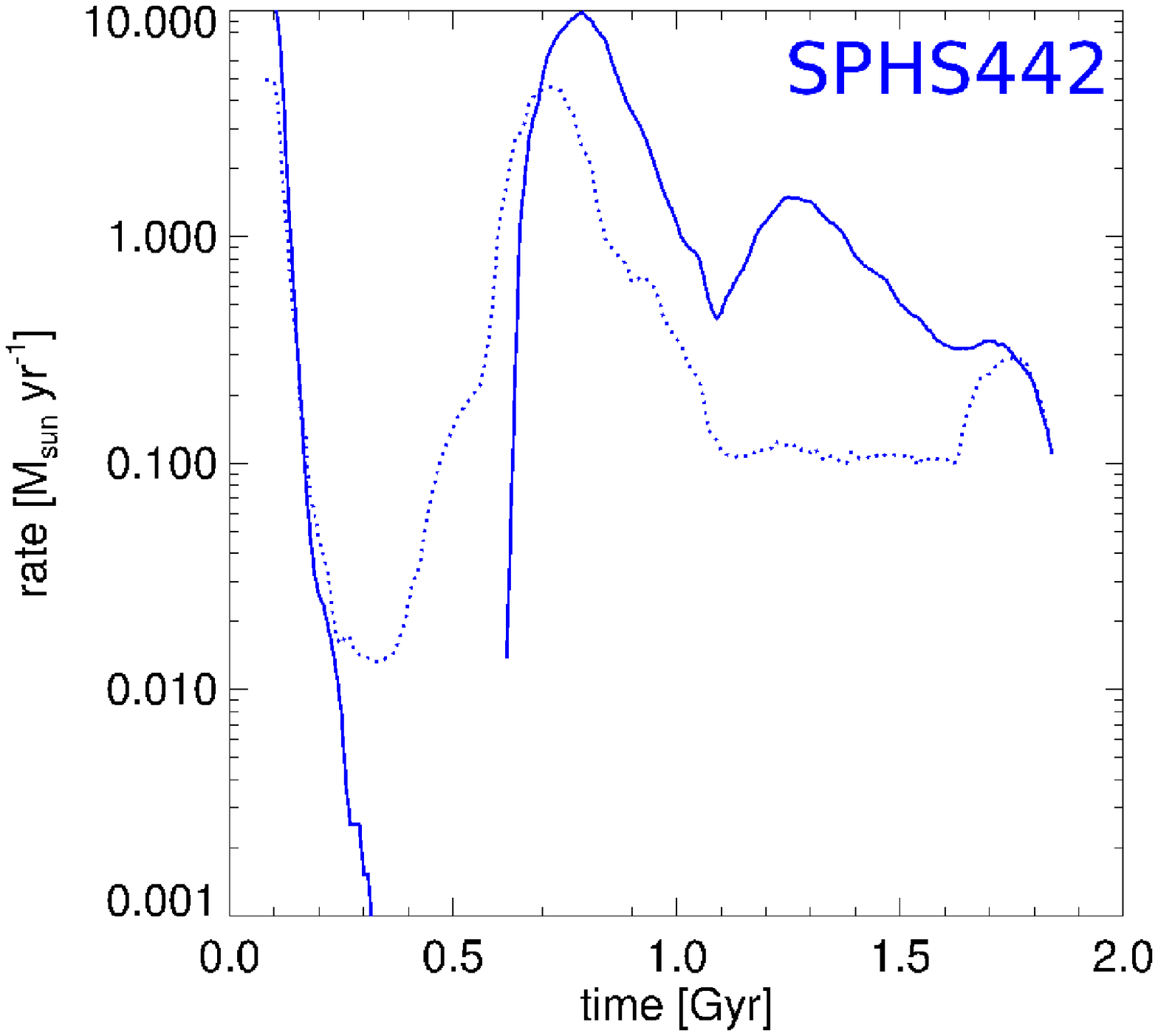,width=1.0\textwidth,angle=0}}
\end{minipage}
\caption[]{Projected surface density plot of the SPHS-442-res2 run at $t = 0.73$\,Gyr. As in the lower resolution case, we see the formation of one or more cold filaments that have formed out of the surrounding gas as a result of the interaction of supernovae-driven hot bubbles. The filaments are funnelling gas down onto the disc at the centre of the computational domain. The progenitors of gas-rich clumps that are formed in the filaments and in the disc are marked in blue.}
\label{fig:streamclumps}
\end{figure}

Here, however, we are in a new regime whereby bound clumps are in fact able to form from the hot halo, but only within the filament(s). The latter structure provides a strong density enhancement within which collapse and fragmentation can occur through non-linear thermal instability \citep{JoungEtal2012, FernandezEtal2012}, rather than through a numerical pressure discontinuity as was the case for the clumps in SPH. Figure \ref{fig:streamclumps} shows the state of the inner 30\,kpc at a time when the filaments have recently formed. We used AHF to identify the locations of any self-bound clumps; these are shown in blue. One or two have already collapsed and are forming stars. As can clearly be seen from the Figure, the only locations where clumps arise are in the streams and in the disc at the centre. The overdensity here is $\sim 10-100$, which is consistent with the instability threshold for a non-linear perturbation in \cite{JoungEtal2012}.

The gas flowing through the filaments ends up contributing to the disc, as do the clumps that form from the fragmentation of the filament. Some of these, however, form stars before they can reach the disc, and therefore end up as orbiting stellar clusters, with masses of $\sim 10^7 \msun$. The stellar clusters, while they are still star-forming, contribute to the continued feedback events and can therefore aid the formation of subsequent streams that may feed the disc.

\begin{figure}
\begin{minipage}[b]{.49\textwidth}
\centerline{\psfig{file=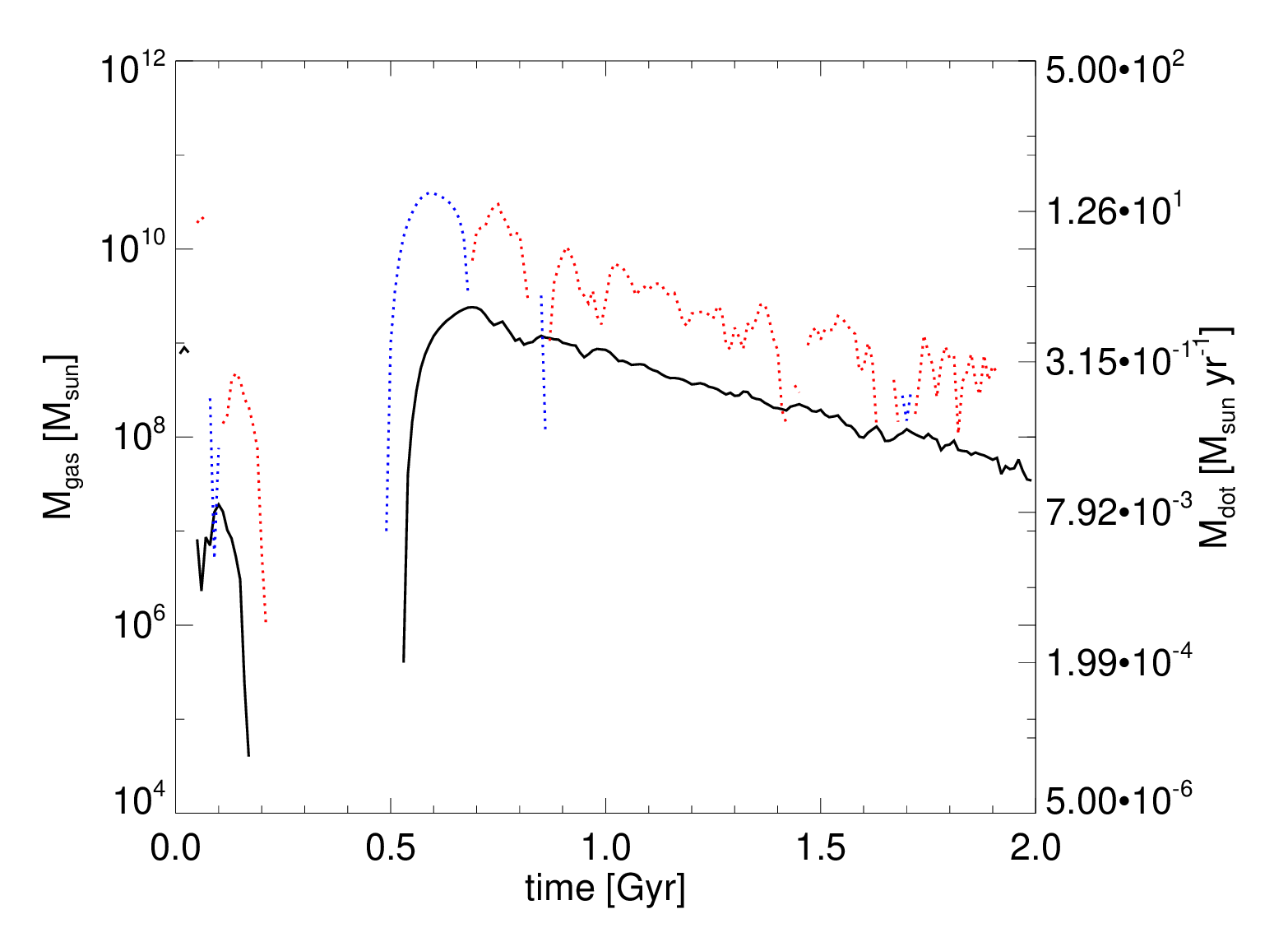,width=1.0\textwidth,angle=0}}
\end{minipage}
\begin{minipage}[b]{.49\textwidth}
\centerline{\psfig{file=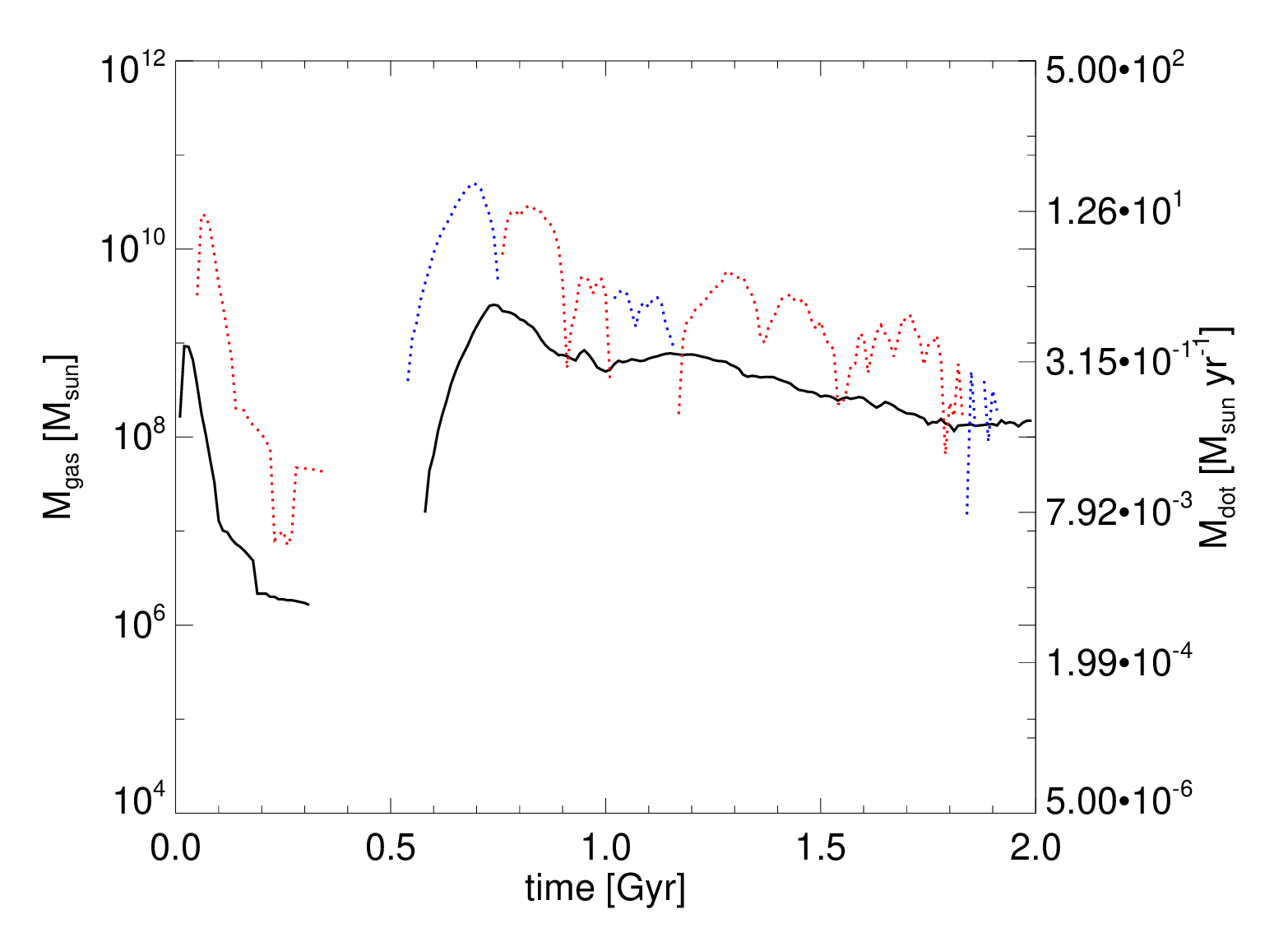,width=1.0\textwidth,angle=0}}
\end{minipage}
\begin{minipage}[b]{.49\textwidth}
\centerline{\psfig{file=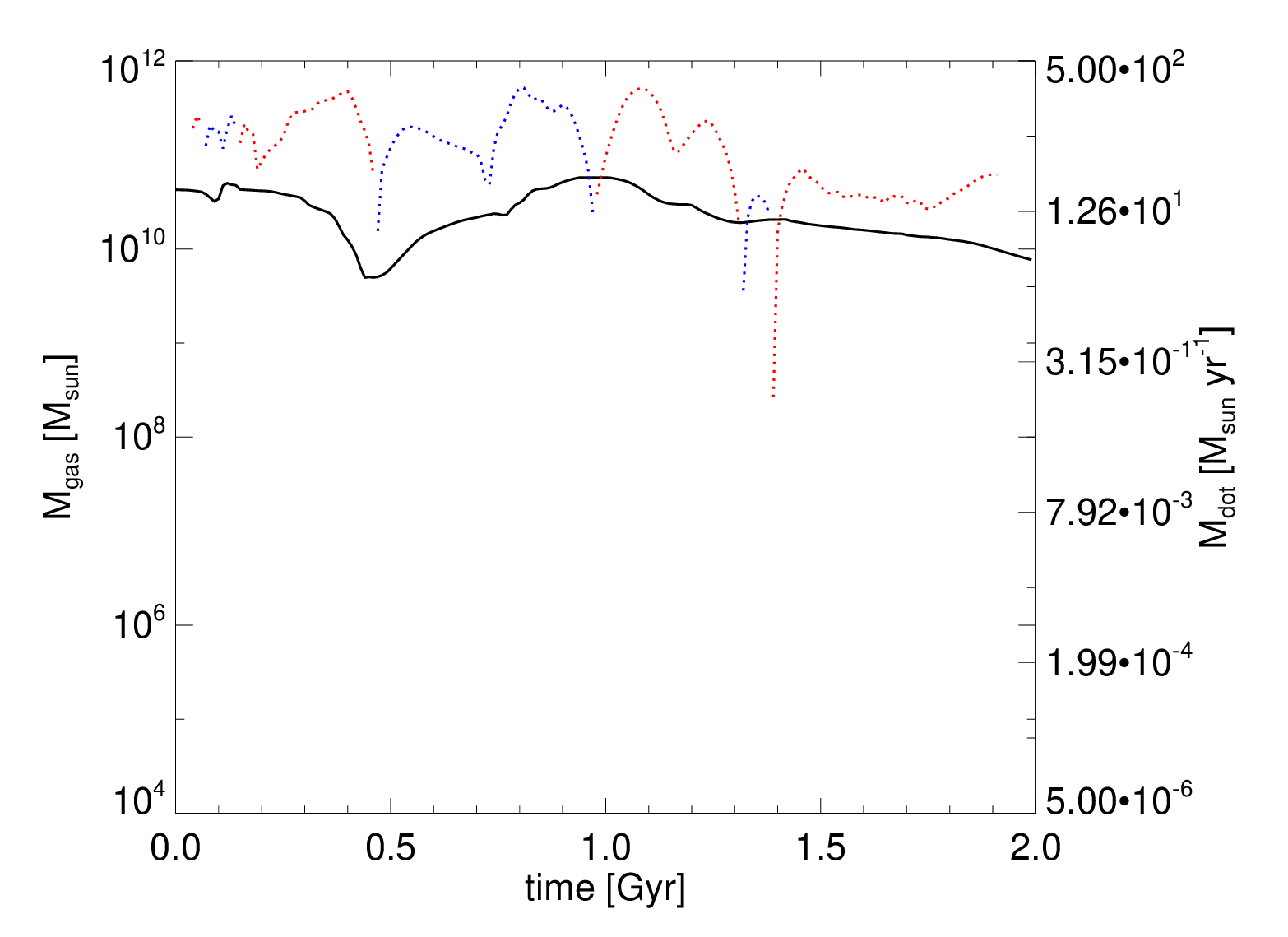,width=1.0\textwidth,angle=0}}
\end{minipage}
\caption[]{Plots of the gas mass contained within a particular region of the phase diagram (black) along with the rate of change of the gas mass into (blue; an `inflow') and out of (red; an `outflow') the phase region. The regions are as per Figure \ref{fig:phasecutsSPHS1.2} but in this case for the SPHS-442-res2 run. The `inflow' and `outflow' rates are plotted in bins of $80$\,Myr. The plots correspond to: (i) the gas in the `isothermal contraction' region (top) - this comprises the temporary disc created in the initial starburst and then later the filaments that form and grow a new disc; (ii) the gas in the `polytrope' region (middle) - this comprises the star forming gas i.e., the clumps that form in the disc and the filaments; and (iii) the gas in the `hot bubble' region, corresponding to gas that has been ejected into the hot halo by supernova feedback.}  
\label{fig:discfeeding}
\end{figure}

In order to quantify the ability of the cold gas in the filaments to grow the disc, in Figure \ref{fig:discfeeding} we plot the gas mass contained in particular phase regions (the same regions marked in Figures \ref{fig:phasecutsSPH1.2} \& \ref{fig:phasecutsSPHS1.2} but for our highest resolution full SPHS run). We also plot the rate of increase or decrease in mass within this region to measure the feeding rate through the filaments. The main period of interest is after the second starburst, and we see from Figure \ref{fig:discfeeding} (top plot) that the streams (and the clumps formed in the streams) are feeding the central disc at a rate of $\sim 1 \msun$ yr$^{-1}$ for the first Gyr after the starburst. This rate is gradually declining over time, and by $t = 2$\,Gyrs has dropped to $\sim 0.1 \msun$ yr$^{-1}$. Looking at the middle plot, we see further that the gas on the polytrope (most of which is in the disc) exhibits a similar rate of decrease in mass; this can be explained by referring to the SFR (bottom) plot in Figure \ref{fig:streamclumps}, which shows that the feeding rate of the disc by the filaments is approximately matched by the star formation rate, since most of the star formation is occurring in the disc at this time.

The 3rd (bottom) plot of Figure \ref{fig:discfeeding} shows the amount (and rate of increase/decrease) of gas in the region of the phase diagram that we have identified as corresponding to the gas recently ejected from supernovae explosions. We see here that this is by far the dominant repository for the gas that has already fallen into the central regions - the mass contained in stars, disc, filaments and clumps combined adds up to $\simlt 1/10$th of the mass that has been ejected into the hot halo, the latter being on average $\sim 1$-$2 \times 10^{10} \msun$.

\section{Discussion}\label{sec:discussion}

We find a clear difference in how cold gas condenses from a hot halo in SPHS versus `classic' SPH. The formation of $\sim$ hundreds of cold clumps from relatively homogeneous regions of the hot halo in SPH owes to a numerical thermal instability. As hot supernovae-driven bubbles collide in the halo, the gas between the bubbles is compressed to high density and cools. The flow is highly shearing and the overdensity rapidly shears away. Due to the lack of multiphase mixing, however, the gas remains cold leading to a pressure dip that seeds the formation of a dense clump of gas. By contrast, in SPHS the shearing gas mixes both in density and entropy and the clumps do not form; instead the gas forms into cold filaments that feed the disc. In our highest resolution SPHS run, these filaments break up to form bound clumps. This fragmentation is physical and owes to a non-linear instability caused by the $\simgt 10$ overdensity in the filament \citep{JoungEtal2012}. 

\cite{KaufmannEtal2006} invoke numerical noise as a potential seed for small-scale fluctuations that could lead to runaway collapse. Indeed, we find that at least for some of the clumps that we see forming in SPH, the seeds are perhaps one or two particles with significantly lower entropies (and initially higher densities) than the gas they end up in as regions of the simulation merge. These lower entropies can lead to pressure dips ($P \propto A \rho^\gamma$, where $A \equiv A(s)$, a function of the specific entropy $s$) when the densities are smoothed across the neighbours, as in SPH the entropy of a particle is `protected' and is not shared. In SPHS this `entropy noise' is effectively removed via the sub-grid diffusion.

There may be a situation where the formation of such clumps from thermal instability is physical, as indeed is suggested by \cite{KaufmannEtal2009} and explored analytically by \cite{BinneyEtal2009}, which is when the entropy profile across the region is nearly completely flat and the cooling timescale can become smaller than the oscillation timescale for small, isotropic perturbations. However, such an entropy profile is highly unlikely for a galaxy formation simulation (both cosmological and non-cosmological) and indeed is certainly not the case in our simulations, where the entropy gradient is quite steep both initially and after multiple supernovae feedback events. Even in this situation, \cite{BinneyEtal2009} find that thermal conduction suppresses linear thermal instabilities through the damping of small-scale perturbations, even for very small fractions of Spitzer's value \citep{Spitzer1962}.

Our finding that at sufficient resolution the SPHS simulations show resolved clumps forming in the overdense filament(s) that feed the disc has particular relevance to the recent work by \cite{FraternaliBinney2008} \& \cite{MarinacciEtal2011}, where the authors suggest a model of gas entrainment from the hot halo through the motion of feedback-driven galactic fountain clouds. These colder clouds mix with the coronal gas and invoke a significant amount of mass transfer from the hot phase to the cold phase, before sinking back again to the galaxy. The key concept here is that the cooling of gas from the halo is dependant on the presence of star-forming gas deeper inside the potential well. Our picture is slightly different, but still within the same vein, as it is the interaction of the walls of multiple supernovae-driven bubbles that cause the condensing of the filaments, which in turn may break up and form cold clumps through non-linear instability as per \cite{JoungEtal2012}. Naturally, therefore, we require that star formation occurs in the disc in order for cold gas to be able to accrete -- a form of {\it positive} feedback. 

Observations by \cite{BoomsmaEtal2005} lend support to this picture, where X-ray emission from hot gas is seen above the plane of the nearby spiral galaxy NGC253, along with an overdense filament-like structure detected in HI at the edge of the hot region, that stretches out for approximately $12$\,kpc. These features are associated with a significant amount of star formation activity and indeed a recent starburst and superwind from the galaxy \citep{HeckmanEtal1990}.

In our highest resolution full-SPHS simulation, we find that the fragmenting filaments formed from overlapping supernovae-driven bubbles feed the disc (and subsequently, star formation in the disc) at a rate of $\sim 1 \msun$ yr$^{-1}$, similar to what is required by observations to fuel star formation in the Milky Way \citep{NohScalo1990, Rocha-PintoEtal2000}. We also find that the gas mass ejected from supernovae explosions remains near-constant around $1$-$2 \times 10^{10} \msun$ for the duration of the simulation ($2$\,Gyrs). If one assumes that the inner hot halo is composed primarily of gas from supernovae feedback processes then this is in excellent agreement with the mass of the hot halo as determined by pulsar dispersion measures \citep[e.g.,][]{2010ApJ...714..320A, GaenslerEtal2008}. We emphasise, however, that our goal in this paper is to study the physics of thermal instabilities in cooling haloes rather than to build a faithful model of Milky Way-mass disc galaxies. Our results are likely sensitive to our sub-grid model parameters, to our resolution, and to our initial conditions that are idealised and non-cosmological. Thus, while we can identify an important thermal instability in the halo that can in principle supply significant cold gas to the disc, it is unclear how important it is with respect to other mechanisms like gas rich mergers, low-redshift cold streams, and gas recycling from stars in the disc. We will explore this in future work.

Finally, a further aspect of the comparison between the SPH and SPHS runs at identical resolution that is particularly interesting is the difference in the kinematic morphologies of the galaxies due to the different cooling mechanisms, i.e., clumps vs. filaments. In the SPH case the cold clumps contributed to a stronger galactic bulge-like feature with a greater amount of material at low angular momentum and on random orbits, while in SPHS the growth of the disc from the cold streams led to a stronger disc feature that was less warped and undergoing a lesser torque from infalling material at different orientations. This result naturally has implications for the so-called `angular momentum problem' in galaxy formation, namely the presence of bulges that are over-dominant and discs that are under-dominant compared to observations \citep[see, e.g.,][]{MallerDekel2002, BurkertDonghia2004, GovernatoEtal2010}.

\section{Conclusions}\label{sec:conclusions}

We have presented simulations of a cooling gaseous halo in a Milky Way mass galaxy, using this particular problem to perform the first scientific investigation with a new hydrodynamics code, SPHS. We have compared the results obtained (at identical spatial resolution) with that of the standard (`classic') SPH method employed in the literature, finding significant differences in the mode of gas cooling in the halo and subsequently the mode of disc feeding. The puzzle of the many cold clumps seen forming from the halo in many SPH simulations of galaxy formation is attributed to a numerical inability driven by unresolved mixing of different gas phases. We demonstrate both with our full simulations (Section \ref{sec:differences}) and with a more idealised test (Section \ref{sec:numericalclump}) that the removal of pressure blips in an otherwise smooth flow prevents the formation of the clumps, leading instead to the formation of cold filaments that feed the disc. The resulting galaxies in the SPH and SPHS simulations differ greatly in their morphology, gas phase diagrams, and stellar and gaseous disc/bulge ratio.

We have explored in more detail the mode of disc feeding seen in our SPHS simulations, going to higher resolution and employing a kernel that gives improved force accuracy. We find a new way of bringing cold gas to the galactic disc; namely, the fragmentation and collapse through non-linear thermal instability of filament(s) formed at the intersection of supernovae-driven bubbles. The feeding rate of cold gas ($T \simlt 10^4$ K) to the disc is found to be approximately a solar mass per year, which suggests this is a promising model for fuelling late-time star formation in real spiral galaxies.

We emphasise that our focus in this paper was on understanding what drives thermal instabilities in hot halo gas, and in particular performing a comparison between the `classic' SPH and the SPHS numerical methods. Our numerical experiments are idealised and do not present a complete picture of galaxy formation. Nonetheless, by employing a hydrodynamics method that resolves the mixing of different gas phases, we find a novel mode of cold gas accretion and disc growth that may be very relevant for galaxy formation.

\section{Acknowledgments}

We thank Volker Springel for the use of {\tt GADGET-3} in this work. We acknowledge useful discussions with Filippo Fraternali, Walter Dehnen, Tobias Kaufmann, and Javiera Guedes. A. Hobbs would like to thank Alexander Knebe for the use of, and guidance on, the Amiga halo finder (AHF). J. I. Read would like to acknowledge support from SNF grant PP00P2\_128540/1. The simulations were performed on the NCi and iVEC facilities at the University of Western Australia (UWA).

\bibliographystyle{mnras}

\bibliography{references}

\begin{thebibliography}{85}
\expandafter\ifx\csname natexlab\endcsname\relax\def\natexlab#1{#1}\fi

\bibitem[{Agertz} et~al.(2007){Agertz}, {Moore}, {Stadel}
  et~al.]{AgertzEtal2007}
{Agertz} O., {Moore} B., {Stadel} J., et~al., 2007, \mnras, 380, 963

\bibitem[{Agertz} et~al.(2009){Agertz}, {Teyssier} \& {Moore}]{AgertzEtal2009}
{Agertz} O., {Teyssier} R., {Moore} B., 2009, \mnras, 397, L64

\bibitem[{Anderson} \& {Bregman}(2010)]{2010ApJ...714..320A}
{Anderson} M.~E., {Bregman} J.~N., 2010, \apj, 714, 320

\bibitem[{Anderson} \& {Bregman}(2011)]{AndersonBregman2011}
{Anderson} M.~E., {Bregman} J.~N., 2011, \apj, 737, 22

\bibitem[{Bate} et~al.(1995){Bate}, {Bonnell} \& {Price}]{Bate95}
{Bate} M.~R., {Bonnell} I.~A., {Price} N.~M., 1995, \mnras, 277, 362

\bibitem[{Bate} \& {Burkert}(1997)]{BateBurkert1997}
{Bate} M.~R., {Burkert} A., 1997, \mnras, 288, 1060

\bibitem[{Binney} et~al.(2009){Binney}, {Nipoti} \&
  {Fraternali}]{BinneyEtal2009}
{Binney} J., {Nipoti} C., {Fraternali} F., 2009, \mnras, 397, 1804

\bibitem[{Boomsma} et~al.(2005){Boomsma}, {Oosterloo}, {Fraternali}, {van der
  Hulst} \& {Sancisi}]{BoomsmaEtal2005}
{Boomsma} R., {Oosterloo} T.~A., {Fraternali} F., {van der Hulst} J.~M.,
  {Sancisi} R., 2005, \aap, 431, 65

\bibitem[{Bullock} et~al.(2001){Bullock}, {Dekel}, {Kolatt}
  et~al.]{BullockEtal2001b}
{Bullock} J.~S., {Dekel} A., {Kolatt} T.~S., et~al., 2001, \apj, 555, 240

\bibitem[{Burkert} \& {D'Onghia}(2004)]{BurkertDonghia2004}
{Burkert} A.~M., {D'Onghia} E., 2004, in { Penetrating Bars Through Masks of
  Cosmic Dust\/}, edited by D.~L. {Block}, I.~{Puerari}, K.~C. {Freeman},
  R.~{Groess}, E.~K. {Block}, vol. 319 of { Astrophysics and Space Science
  Library\/},  341

\bibitem[{Cole} et~al.(2011){Cole}, {Dehnen} \& {Wilkinson}]{ColeEtal2011}
{Cole} D.~R., {Dehnen} W., {Wilkinson} M.~I., 2011, \mnras, 416, 1118

\bibitem[{Collins} et~al.(2005){Collins}, {Shull} \& {Giroux}]{CollinsEtal2005}
{Collins} J.~A., {Shull} J.~M., {Giroux} M.~L., 2005, \apj, 623, 196

\bibitem[{Couchman} et~al.(1995){Couchman}, {Thomas} \&
  {Pearce}]{1995ApJ...452..797C}
{Couchman} H.~M.~P., {Thomas} P.~A., {Pearce} F.~R., 1995, \apj, 452, 797

\bibitem[{Cullen} \& {Dehnen}(2010)]{2010MNRAS.408..669C}
{Cullen} L., {Dehnen} W., 2010, \mnras, 408, 669

\bibitem[{Dehnen} \& {Aly}(2012)]{DehnenAly2012}
{Dehnen} W., {Aly} H., 2012, ArXiv e-prints

\bibitem[{Dehnen} \& {McLaughlin}(2005)]{DehnenMcLaughlin2005}
{Dehnen} W., {McLaughlin} D.~E., 2005, \mnras, 363, 1057

\bibitem[{Dobbs} et~al.(2011){Dobbs}, {Burkert} \& {Pringle}]{DobbsEtal2011}
{Dobbs} C.~L., {Burkert} A., {Pringle} J.~E., 2011, \mnras, 413, 2935

\bibitem[{Dubois} et~al.(2012){Dubois}, {Devriendt}, {Slyz} \&
  {Teyssier}]{DuboisEtal2012}
{Dubois} Y., {Devriendt} J., {Slyz} A., {Teyssier} R., 2012, \mnras, 420, 2662

\bibitem[{Fang} et~al.(2006){Fang}, {Mckee}, {Canizares} \&
  {Wolfire}]{FangEtal2006}
{Fang} T., {Mckee} C.~F., {Canizares} C.~R., {Wolfire} M., 2006, \apj, 644, 174

\bibitem[{Fern{\'a}ndez} et~al.(2012){Fern{\'a}ndez}, {Joung} \&
  {Putman}]{FernandezEtal2012}
{Fern{\'a}ndez} X., {Joung} M.~R., {Putman} M.~E., 2012, \apj, 749, 181

\bibitem[{Fraternali} \& {Binney}(2008)]{FraternaliBinney2008}
{Fraternali} F., {Binney} J.~J., 2008, \mnras, 386, 935

\bibitem[{Fraternali} \& {Tomassetti}(2012)]{FraternaliTomassetti2012}
{Fraternali} F., {Tomassetti} M., 2012, ArXiv e-prints

\bibitem[{Fukugita} \& {Peebles}(2006)]{FukugitaPeebles2006}
{Fukugita} M., {Peebles} P.~J.~E., 2006, \apj, 639, 590

\bibitem[{Gaensler} et~al.(2008){Gaensler}, {Madsen}, {Chatterjee} \&
  {Mao}]{GaenslerEtal2008}
{Gaensler} B.~M., {Madsen} G.~J., {Chatterjee} S., {Mao} S.~A., 2008, \pasa,
  25, 184

\bibitem[{Gill} et~al.(2004){Gill}, {Knebe} \& {Gibson}]{AHF1}
{Gill} S.~P.~D., {Knebe} A., {Gibson} B.~K., 2004, \mnras, 351, 399

\bibitem[{Governato} et~al.(2010){Governato}, {Brook}, {Mayer}
  et~al.]{GovernatoEtal2010}
{Governato} F., {Brook} C., {Mayer} L., et~al., 2010, \nat, 463, 203

\bibitem[{Grcevich} \& {Putman}(2009)]{GrcevichPutman2009}
{Grcevich} J., {Putman} M.~E., 2009, \apj, 696, 385

\bibitem[{Gupta} et~al.(2012){Gupta}, {Mathur}, {Krongold}, {Nicastro} \&
  {Galeazzi}]{GuptaEtal2012}
{Gupta} A., {Mathur} S., {Krongold} Y., {Nicastro} F., {Galeazzi} M., 2012,
  ArXiv e-prints

\bibitem[{Heckman} et~al.(1990){Heckman}, {Armus} \& {Miley}]{HeckmanEtal1990}
{Heckman} T.~M., {Armus} L., {Miley} G.~K., 1990, \apjs, 74, 833

\bibitem[{Heitsch} \& {Putman}(2009)]{HeitschPutman2009}
{Heitsch} F., {Putman} M.~E., 2009, \apj, 698, 1485

\bibitem[{Hippelein} et~al.(2003){Hippelein}, {Maier}, {Meisenheimer}
  et~al.]{HippeleinEtal2003}
{Hippelein} H., {Maier} C., {Meisenheimer} K., et~al., 2003, \aap, 402, 65

\bibitem[{Joung} et~al.(2012){Joung}, {Bryan} \& {Putman}]{JoungEtal2012}
{Joung} M.~R., {Bryan} G.~L., {Putman} M.~E., 2012, \apj, 745, 148

\bibitem[{Kacprzak} et~al.(2008){Kacprzak}, {Churchill}, {Steidel} \&
  {Murphy}]{KacprzakEtal2008}
{Kacprzak} G.~G., {Churchill} C.~W., {Steidel} C.~C., {Murphy} M.~T., 2008,
  \aj, 135, 922

\bibitem[{Katz} et~al.(1996){Katz}, {Weinberg} \& {Hernquist}]{KWH1996}
{Katz} N., {Weinberg} D.~H., {Hernquist} L., 1996, \apjs, 105, 19

\bibitem[{Kaufmann} et~al.(2009){Kaufmann}, {Bullock}, {Maller}, {Fang} \&
  {Wadsley}]{KaufmannEtal2009}
{Kaufmann} T., {Bullock} J.~S., {Maller} A.~H., {Fang} T., {Wadsley} J., 2009,
  \mnras, 396, 191

\bibitem[{Kaufmann} et~al.(2006){Kaufmann}, {Mayer}, {Wadsley}, {Stadel} \&
  {Moore}]{KaufmannEtal2006}
{Kaufmann} T., {Mayer} L., {Wadsley} J., {Stadel} J., {Moore} B., 2006, \mnras,
  370, 1612

\bibitem[{Kaufmann} et~al.(2007){Kaufmann}, {Mayer}, {Wadsley}, {Stadel} \&
  {Moore}]{KaufmannEtal2007}
{Kaufmann} T., {Mayer} L., {Wadsley} J., {Stadel} J., {Moore} B., 2007, \mnras,
  375, 53

\bibitem[{Kennicutt} et~al.(1994){Kennicutt}, {Tamblyn} \&
  {Congdon}]{KennicuttEtal1994}
{Kennicutt} Jr. R.~C., {Tamblyn} P., {Congdon} C.~E., 1994, \apj, 435, 22

\bibitem[{Knollmann} \& {Knebe}(2009)]{AHF2}
{Knollmann} S.~R., {Knebe} A., 2009, \apjs, 182, 608

\bibitem[{Lada} \& {Lada}(2003)]{LadaLada2003}
{Lada} C.~J., {Lada} E.~A., 2003, \araa, 41, 57

\bibitem[{Leitner} \& {Kravtsov}(2011)]{LeitnerKravtsov2011}
{Leitner} S.~N., {Kravtsov} A.~V., 2011, \apj, 734, 48

\bibitem[{Lilly} et~al.(1996){Lilly}, {Le Fevre}, {Hammer} \&
  {Crampton}]{LillyEtal1996}
{Lilly} S.~J., {Le Fevre} O., {Hammer} F., {Crampton} D., 1996, \apjl, 460, L1

\bibitem[{Madau} et~al.(1998){Madau}, {Pozzetti} \& {Dickinson}]{MadauEtal1998}
{Madau} P., {Pozzetti} L., {Dickinson} M., 1998, \apj, 498, 106

\bibitem[{Malagoli} et~al.(1987){Malagoli}, {Rosner} \&
  {Bodo}]{MalagoliEtal1987}
{Malagoli} A., {Rosner} R., {Bodo} G., 1987, \apj, 319, 632

\bibitem[{Maller} \& {Dekel}(2002)]{MallerDekel2002}
{Maller} A.~H., {Dekel} A., 2002, \mnras, 335, 487

\bibitem[{Marasco} et~al.(2012){Marasco}, {Fraternali} \&
  {Binney}]{MarascoEtal2012}
{Marasco} A., {Fraternali} F., {Binney} J.~J., 2012, \mnras, 419, 1107

\bibitem[{Marinacci} et~al.(2011){Marinacci}, {Fraternali}, {Nipoti}, {Binney},
  {Ciotti} \& {Londrillo}]{MarinacciEtal2011}
{Marinacci} F., {Fraternali} F., {Nipoti} C., {Binney} J., {Ciotti} L.,
  {Londrillo} P., 2011, \mnras, 415, 1534

\bibitem[{Maron} \& {Howes}(2003)]{2003ApJ...595..564M}
{Maron} J.~L., {Howes} G.~G., 2003, \apj, 595, 564

\bibitem[{Mashchenko} et~al.(2008){Mashchenko}, {Wadsley} \&
  {Couchman}]{MashchenkoEtal2008}
{Mashchenko} S., {Wadsley} J., {Couchman} H.~M.~P., 2008, Science, 319, 174

\bibitem[{Mastropietro} et~al.(2005){Mastropietro}, {Moore}, {Mayer}, {Wadsley}
  \& {Stadel}]{MastropietroEtal2005}
{Mastropietro} C., {Moore} B., {Mayer} L., {Wadsley} J., {Stadel} J., 2005,
  \mnras, 363, 509

\bibitem[{Mo} et~al.(2005){Mo}, {Yang}, {van den Bosch} \& {Katz}]{MoEtal2005}
{Mo} H.~J., {Yang} X., {van den Bosch} F.~C., {Katz} N., 2005, \mnras, 363,
  1155

\bibitem[{Navarro} et~al.(1996){Navarro}, {Frenk} \&
  {White}]{NavarroFrenkWhite1996}
{Navarro} J.~F., {Frenk} C.~S., {White} S.~D.~M., 1996, \apj, 462, 563

\bibitem[{Nicastro} et~al.(2008){Nicastro}, {Mathur} \&
  {Elvis}]{NicastroEtal2008}
{Nicastro} F., {Mathur} S., {Elvis} M., 2008, Science, 319, 55

\bibitem[{Noh} \& {Scalo}(1990)]{NohScalo1990}
{Noh} H.-R., {Scalo} J., 1990, \apj, 352, 605

\bibitem[{Nulsen}(1986)]{Nulsen1986}
{Nulsen} P.~E.~J., 1986, \mnras, 221, 377

\bibitem[{Peek}(2009)]{Peek2009}
{Peek} J.~E.~G., 2009, \apj, 698, 1429

\bibitem[{Price}(2008)]{2008JCoPh.22710040P}
{Price} D.~J., 2008, Journal of Computational Physics, 227, 10040

\bibitem[{Price} \& {Monaghan}(2007)]{Price2007}
{Price} D.~J., {Monaghan} J.~J., 2007, \mnras, 374, 1347

\bibitem[{Putman} et~al.(2009){Putman}, {Peek} \& {Heitsch}]{PutmanEtal2009}
{Putman} M.~E., {Peek} J.~E.~G., {Heitsch} F., 2009, ArXiv e-prints

\bibitem[{Read} \& {Hayfield}(2012)]{2012MNRAS.tmp.2941R}
{Read} J.~I., {Hayfield} T., 2012, \mnras,  2941

\bibitem[{Read} et~al.(2010{\natexlab{a}}){Read}, {Hayfield} \&
  {Agertz}]{ReadEtal2010}
{Read} J.~I., {Hayfield} T., {Agertz} O., 2010{\natexlab{a}}, \mnras, 405, 1513

\bibitem[{Read} et~al.(2010{\natexlab{b}}){Read}, {Hayfield} \&
  {Agertz}]{2009arXiv0906.0774R}
{Read} J.~I., {Hayfield} T., {Agertz} O., 2010{\natexlab{b}}, \mnras, 405, 1513

\bibitem[{Read} \& {Trentham}(2005)]{2005RSPTA.363.2693R}
{Read} J.~I., {Trentham} N., 2005, Royal Society of London Philosophical
  Transactions Series A, 363, 2693

\bibitem[{Roberts}(1963)]{Roberts1963}
{Roberts} M.~S., 1963, \araa, 1, 149

\bibitem[{Rocha-Pinto} et~al.(2000){Rocha-Pinto}, {Scalo}, {Maciel} \&
  {Flynn}]{Rocha-PintoEtal2000}
{Rocha-Pinto} H.~J., {Scalo} J., {Maciel} W.~J., {Flynn} C., 2000, \apjl, 531,
  L115

\bibitem[{Saitoh} \& {Makino}(2009)]{SaitohMakino2009}
{Saitoh} T.~R., {Makino} J., 2009, \apjl, 697, L99

\bibitem[{Salpeter}(1955)]{Salpeter1955}
{Salpeter} E.~E., 1955, \apj, 121, 161

\bibitem[{Sancisi} et~al.(2008){Sancisi}, {Fraternali}, {Oosterloo} \& {van der
  Hulst}]{SancisiEtal2008}
{Sancisi} R., {Fraternali} F., {Oosterloo} T., {van der Hulst} T., 2008, \aapr,
  15, 189

\bibitem[{Sandage}(1986)]{Sandage1986}
{Sandage} A., 1986, \aap, 161, 89

\bibitem[{Schmidt}(1959)]{Schmidt1959}
{Schmidt} M., 1959, \apj, 129, 243

\bibitem[{Sedov}(1959)]{Sedov1959}
{Sedov} L.~I., 1959, {Similarity and Dimensional Methods in Mechanics}

\bibitem[{Sembach} et~al.(2003){Sembach}, {Wakker}, {Savage}
  et~al.]{SembachEtal2003}
{Sembach} K.~R., {Wakker} B.~P., {Savage} B.~D., et~al., 2003, \apjs, 146, 165

\bibitem[{Sommer-Larsen}(2006)]{Sommer-Larsen2006}
{Sommer-Larsen} J., 2006, \apjl, 644, L1

\bibitem[{Spitzer}(1962)]{Spitzer1962}
{Spitzer} L., 1962, {Physics of Fully Ionized Gases}

\bibitem[{Springel}(2005)]{Springel2005}
{Springel} V., 2005, \mnras, 364, 1105

\bibitem[{Springel} \& {Hernquist}(2002{\natexlab{a}})]{SpringelHernquist02}
{Springel} V., {Hernquist} L., 2002{\natexlab{a}}, \mnras, 333, 649

\bibitem[{Springel} \& {Hernquist}(2002{\natexlab{b}})]{SpringelHernquist2002}
{Springel} V., {Hernquist} L., 2002{\natexlab{b}}, \mnras, 333, 649

\bibitem[{Stewart} et~al.(2011){Stewart}, {Kaufmann}, {Bullock}
  et~al.]{StewartEtal2011}
{Stewart} K.~R., {Kaufmann} T., {Bullock} J.~S., et~al., 2011, \apjl, 735, L1

\bibitem[{Tripp} et~al.(2003){Tripp}, {Wakker}, {Jenkins}
  et~al.]{TrippEtal2003}
{Tripp} T.~M., {Wakker} B.~P., {Jenkins} E.~B., et~al., 2003, \aj, 125, 3122

\bibitem[{Truelove} et~al.(1997){Truelove}, {Klein}, {McKee}, {Holliman},
  {Howell} \& {Greenough}]{TrueloveEtal1997}
{Truelove} J.~K., {Klein} R.~I., {McKee} C.~F., {Holliman} II J.~H., {Howell}
  L.~H., {Greenough} J.~A., 1997, \apjl, 489, L179

\bibitem[{van de Voort} et~al.(2011){van de Voort}, {Schaye}, {Booth}, {Haas}
  \& {Dalla Vecchia}]{vandeVoortEtal2011}
{van de Voort} F., {Schaye} J., {Booth} C.~M., {Haas} M.~R., {Dalla Vecchia}
  C., 2011, \mnras, 414, 2458

\bibitem[{Vogelsberger} et~al.(2011){Vogelsberger}, {Sijacki}, {Keres},
  {Springel} \& {Hernquist}]{VogelsbergerEtal2011}
{Vogelsberger} M., {Sijacki} D., {Keres} D., {Springel} V., {Hernquist} L.,
  2011, ArXiv e-prints

\bibitem[{Wadsley} et~al.(2004){Wadsley}, {Stadel} \&
  {Quinn}]{2004NewA....9..137W}
{Wadsley} J.~W., {Stadel} J., {Quinn} T., 2004, New Astronomy, 9, 137

\bibitem[{Wadsley} et~al.(2008){Wadsley}, {Veeravalli} \&
  {Couchman}]{2008MNRAS.387..427W}
{Wadsley} J.~W., {Veeravalli} G., {Couchman} H.~M.~P., 2008, \mnras, 387, 427

\bibitem[{Williams} et~al.(2005){Williams}, {Mathur}, {Nicastro}
  et~al.]{WilliamsEtal2005}
{Williams} R.~J., {Mathur} S., {Nicastro} F., et~al., 2005, \apj, 631, 856

\end{thebibliography}

\end{document}